\documentclass[fleqn,usenatbib]{mnras}

\usepackage[T1]{fontenc}

\DeclareRobustCommand{\VAN}[3]{#2}
\let\VANthebibliography\thebibliography
\def\thebibliography{\DeclareRobustCommand{\VAN}[3]{##3}\VANthebibliography}

\usepackage{graphicx}	
\usepackage{amsmath}	
\usepackage{amssymb}	
\usepackage{multirow}

\usepackage{subfig}

\usepackage{xcolor}
\usepackage{bm}

\usepackage{anyfontsize}

\newcommand{\Dg}{\Delta_{\rm g}}
\newcommand{\bg}{b_{\rm lin}}
\newcommand{\bq}{b_{\rm mag}}
\newcommand{\be}{b_{\rm evo}}
\newcommand{\cH}{\mathcal{H}}

\newcommand{\de}{{\rm d}}

\newcommand{\cP}{\mathcal{P}}
\newcommand{\cW}{\mathcal{W}}

\newcommand{\cex}{C^{\rm Ex}_\ell}
\newcommand{\capp}{C^{\rm Ap}_\ell}
\newcommand{\cijex}{C^{\rm Ex}_{ij\ell}}
\newcommand{\cijapp}{C^{\rm Ap}_{ij\ell}}

\title[]{Ultra-large-scale approximations and galaxy clustering: debiasing constraints on cosmological parameters}

\author[Martinelli et al.]{
\parbox{\textwidth}{
Matteo\
Martinelli,$^1$\thanks{\href{mailto:matteo.martinelli@uam.es}{matteo.martinelli@uam.es}}
Roohi\
Dalal,$^2$\thanks{\href{mailto:rdalal@princeton.edu}{rdalal@princeton.edu}}
Fereshteh\
Majidi,$^{3,4}$\thanks{\href{mailto:fereshteh.majidi@gmail.com}{fereshteh.majidi@gmail.com}}
Yashar\
Akrami,$^{5,6}$\thanks{\href{mailto:akrami@ens.fr}{akrami@ens.fr}}\\
Stefano\
Camera,$^{7,8,9,10}$\thanks{\href{mailto:stefano.camera@unito.it}{stefano.camera@unito.it} (\texttt{ORCID: 0000-0003-3399-3574})}
and Elena\
Sellentin$^{11}$\thanks{\href{mailto:sellentin@strw.leidenuniv.nl}{sellentin@strw.leidenuniv.nl}}}
\vspace{0.4cm}
\\
$^1$Instituto de F\'isica Te\'orica, Universidad Aut\'onoma de Madrid, Campus de Cantoblanco, 28049 Madrid, Spain\\
$^2$Department of Astrophysical Sciences, Princeton University, Peyton Hall, Princeton, NJ 08544, USA\\
$^3$Department of Physics and Chemistry, Alzahra University, Vanak Village Street, Tehran, Iran\\
$^4$Department of Physics and Astronomy, University of British Columbia, 6224 Agricultural Road, Vancouver, B.C., Canada\\
$^{5}$Laboratoire de Physique de l'\'Ecole Normale Sup\'erieure, Universite PSL, CNRS, Sorbonne Universit\'e, 75005 Paris, France\\
$^{6}$Observatoire de Paris, Universit\'e PSL, Sorbonne Universit\'e, LERMA, 75014 Paris, France\\
$^7$Dipartimento di Fisica, Universit\`a degli Studi di Torino, via P.\ Giuria 1, 10125 Torino, Italy\\
$^8$INFN -- Istituto Nazionale di Fisica Nucleare, Sezione di Torino, via P.\ Giuria 1, 10125 Torino, Italy\\
$^9$INAF -- Istituto Nazionale di Astrofisica, Osservatorio Astrofisico di Torino, strada Osservatorio 20, 10025 Pino Torinese, Italy\\
$^{10}$Department of Physics \& Astronomy, University of the Western Cape, Cape Town 7535, South Africa\\
$^{11}$Leiden Observatory, Leiden University, Huygens Laboratory, Niels Bohrweg 2, NL-2333 CA, Leiden, The Netherlands
}

\date{Accepted 2021 December 3. Received 2021 November 23; in original form 2021 August 13}

\pubyear{2021}

\begin{document}
\label{firstpage}
\pagerange{\pageref{firstpage}--\pageref{lastpage}}

\maketitle

\begin{abstract}
Upcoming galaxy surveys will allow us to probe the growth of the cosmic large-scale structure with improved sensitivity compared to current missions, and will also map larger areas of the sky. This means that in addition to the increased precision in observations, future surveys will also access the ultra-large-scale regime, where commonly neglected effects such as lensing, redshift-space distortions and relativistic corrections become important for calculating correlation functions of galaxy positions. At the same time, several approximations usually made in these calculations, such as the Limber approximation, break down at those scales. The need to abandon these approximations and simplifying assumptions at large scales creates severe issues for parameter estimation methods. On the one hand, exact calculations of theoretical angular power spectra become computationally expensive, and the need to perform them thousands of times to reconstruct posterior probability distributions for cosmological parameters makes the approach unfeasible. On the other hand, neglecting relativistic effects and relying on approximations may significantly bias the estimates of cosmological parameters. In this work, we quantify this bias and investigate how an incomplete modelling of various effects on ultra-large scales could lead to false detections of new physics beyond the standard $\Lambda$CDM model. Furthermore, we propose a simple debiasing method that allows us to recover true cosmologies without running the full parameter estimation pipeline with exact theoretical calculations. This method can therefore provide a fast way of obtaining accurate values of cosmological parameters and estimates of exact posterior probability distributions from ultra-large-scale observations.
\end{abstract}




\section{Introduction}\label{sec:intro}

In recent years, the development of cosmic microwave background observations, led by surveys such as the Wilkinson Microwave Anisotropy Probe (WMAP) \citep{2013ApJS..208...19H}, {\it Planck} \citep{2020A&A...641A...1P,2020A&A...641A...6P}, the South Pole Telescope (SPT) \citep{Carlstrom:2009um} and the Atacama Cosmology Telescope (ACT) \citep{Aiola:2020azj}, has brought cosmology into the precision era. The new frontier for cosmological observations is to now reach a similar precision in surveys of the cosmic large-scale structure. Observations of the large-scale structure can provide information on the matter distribution in the Universe and on the growth of primordial perturbations with time. This is achieved, for example, by observing the lensing effect of intervening matter on background galaxies (cosmic shear) or by measuring the correlation function of the positions of galaxies (galaxy clustering). The former has been the main focus of the Kilo-Degree Survey (KiDS) collaboration which has provided constraints on cosmological parameters both for the standard $\Lambda$CDM model and for some extensions \citep{Kohlinger:2017sxk}. The latter has been explored to exquisite precision by several observational collaborations such as the two-degree Field Galaxy Redshift Survey \citep{Cole:2005sx}, the six-degree Field Galaxy Survey \citep{2011MNRAS.416.3017B}, WiggleZ \citep{2011MNRAS.418.1707B,2012PhRvD..86j3518P} and the Sloan Digital Sky Survey (SDSS) \citep{Eisenstein:2005su,2010MNRAS.401.2148P,2012MNRAS.427.3435A,Alam:2016hwk}. Experiments like the Dark Energy Survey (DES) have recently provided state-of-the-art measurements of cosmological parameters using both shear and clustering from photometric measurements \citep{Abbott:2021bzy}.

In the near future, observations of the large-scale structure will be further improved by new missions, either space-borne such as {\it Euclid} \citep{2011arXiv1110.3193L,2013LRR....16....6A,2018LRR....21....2A,2020A&A...642A.191E}, the Roman Space Telescope \citep{2015arXiv150303757S} and the Spectro-Photometer for the History of the Universe, Epoch of Reionization, and Ices Explorer (SPHEREx) \citep{2014arXiv1412.4872D,2018arXiv180505489D}, or ground-based such as the Dark Energy Spectroscopic Instrument (DESI) \citep{2016arXiv161100036D,2016arXiv161100037D}, the Rubin Observatory Legacy Survey of Space and Time (LSST) \citep{2009arXiv0912.0201L,2018arXiv180901669T,Ivezic:2008fe} and the SKA Observatory (SKAO) \citep{2015aska.confE..17A,2015aska.confE..19S,2015aska.confE..23B,2015aska.confE..24B,2015aska.confE..25C,2015aska.confE..31R,2020PASA...37....7S}. These future surveys will indeed improve the sensitivity of the measurements, and, in addition, will make it possible to perform observations on large volumes of the sky. With such observations, it will be possible to access, for the first time, ultra-large scales when measuring the correlation function of galaxy positions and shear. While this ability to access such large scales will allow us to better constrain cosmological models and test fundamental theories such as general relativity \citep{2015ApJ...811..116B,2021arXiv210512582S}, it will also pose new challenges to our ability to theoretically model the observables involved.

In particular, the galaxy correlation function at very large scales receives contributions from lensing, redshift-space distortions (RSD) and relativistic effects \citep{Yoo:2010ni,Bonvin:2011bg,Challinor:2011bk,2014JCAP...09..037B}, which are mostly negligible for the scales probed by current surveys \citep[see e.g.][]{2015MNRAS.447.1789Y,Fonseca:2015laa,2015ApJ...814..145A}. The modelling problem presented by such contributions is not as severe as the one of modelling nonlinear effects at small scales, where one needs to rely on model-dependent numerical simulations \citep[see e.g.][]{Martinelli:2020yto,Safi:2020ugb,Bose:2021mkz,2021MNRAS.503.1897C,2021arXiv210611718C}. However, in order to simplify the modelling of large-scale effects, several approximations are commonly made in computing theoretical predictions for galaxy number counts, such as the Limber \citep{LoVerde:2008re} and the flat-sky \citep{Matthewson:2020rdt} approximations. Such simplifications hold for the scales probed by current surveys \citep{Kilbinger:2017lvu}, but they may fail when larger scales will be accessed by future surveys.

Calculations that include large-scale effects and do not rely on approximations are feasible, and codes commonly used to compute theoretical predictions, such as \texttt{CAMB} \citep{Lewis:1999bs,Howlett:2012mh} and \texttt{CLASS} \citep{2011JCAP...07..034B}, allow us to obtain `exact' galaxy clustering power spectra. However, the computational time required for such exact calculations is significantly longer, causing parameter estimation pipelines to become unfeasible, as they require calculating tens of thousands of spectra to reconstruct posterior probability distributions for cosmological parameters.

Several attempts have been made to overcome this problem. For instance, fast Fourier transform (FFT) or logarithmic FFT (FFTLog) methods can be exploited to accelerate the computation of the theoretical predictions \citep{2017JCAP...11..054A,2017A&A...602A..72C,2018PhRvD..97b3504G}. Alternatively, approximations can be made to reduce the dimensionality of the integration, namely either assuming that the observed patch of sky is flat, and thus performing a two-dimensional Fourier transform on the sky \citep{Datta:2006vh,White:2017ufc,Jalilvand:2019brk,Matthewson:2020rdt}, or exploiting the behaviour of spherical Bessel functions at large angular multipoles \citep{1953ApJ...117..134L,1954ApJ...119..655L,1992ApJ...388..272K}.

In this work, we investigate how applying these commonly used approximations and neglecting lensing, RSD and relativistic contributions at large scales can bias the estimation of cosmological parameters, and possibly lead to false detections of non-standard cosmological models. Such an analysis has been of interest for some time \citep[see e.g.][]{Camera:2014dia,Camera:2014sba,Thiele:2019fcu,Villa:2017yfg}, but we investigate it here considering all the large-scale effects and approximations at the same time, while relying on a full Markov chain Monte Carlo (MCMC) pipeline for parameter estimation, rather than using Fisher matrices. Note that other studies \citep[e.g.][]{cardona2016,Tanidis:2019teo,Tanidis:2019fdh} did approach the problem from the MCMC point of view, but they all, in one way or another, had to simplify the problem in a way that either made them differ from a benchmark analysis, or assumed some of the aforementioned approximations.

Additionally, we propose a simple debiasing method to recover the true values of cosmological parameters without the need for exact calculations of the power spectra. Such a method will allow us to analyse future data sets in a manner that avoids computational problems, but ensures that we accurately obtain the correct best-fit values of cosmological parameters and estimates of their posterior distributions.

The paper is structured as follows. We review in \autoref{sec:GC} the theoretical modelling of galaxy number count correlations, presenting both the exact computation and the approximated one. In \autoref{sec:survey}, the experimental setup used throughout the paper is presented, while in \autoref{sec:cases} we describe the cosmological models considered in this paper and their impacts on galaxy number counts. In \autoref{sec:methods}, we present our analysis pipeline and introduce a debiasing method able to significantly reduce the bias on cosmological parameters introduced by incorrect modelling of the observables. We present our results in \autoref{sec:results} and draw our conclusions in \autoref{sec:conclusions}. 

\section{Galaxy number counts and  harmonic-space correlation functions}\label{sec:GC}
Observed fluctuations in galaxy number counts are primarily caused by underlying inhomogeneities in the matter density field on cosmological scales and, for galaxies, are a biased tracer of the cosmic large-scale structure. However, there is a score of secondary effects that also contribute to the observed signal \citep{Yoo:2010ni,Challinor:2011bk,Bonvin:2011bg}.  The most important of them are the well-known redshift-space distortions, which represent the dominant term on sub-Hubble scales, and weak lensing magnification, important for deep surveys and wide redshift bins. Additionally, there is a more complicated set of relativistic terms that arise from radial and transverse perturbations along the photon path from the source to the observer.

Thus, we can write the observed galaxy number count fluctuation field in real space and up to first order in cosmological perturbation theory as \citep[see e.g.][]{2018JCAP...06..008G}\footnote{\label{Footnote1}Note that several different symbols are used in the literature to denote the magnification bias and---as we shall see later on---the evolution bias, e.g.\ \(\alpha\), \(\mathcal Q\), and \(s\) for the former, and \(b_{\rm e}\) and \(f_{\rm evo}\) for the latter \citep[see also][]{2021arXiv210713401M}. Here, however, we adopt a more uniform notation, with \(\bg\), \(\bq\), and \(\be\) respectively denoting the linear galaxy bias, the magnification bias, and the evolution bias. For the first two, the rationale behind our notation is that they respectively are what modulates the matter density fluctuations and lensing convergence.}
\begin{equation}
    \Dg=
    \bg\,\delta
    -\frac{1}{\cH}\partial_\parallel^2V
    -
    \bq\,\kappa
    +\Delta_{\rm loc}+\Delta_{\rm int}\;.\label{eq:Delta_g}
\end{equation}
(Note that hereafter we shall use units such that $c=1$.) To understand better what the expression above means, we shall now break it up in all its terms:
\begin{enumerate}
    \item The first term in \autoref{eq:Delta_g} sees the linear galaxy bias, \(\bg\), multiplying matter density fluctuations in the comoving-synchronous gauge, \(\delta
    \).
    \item The second term is linear RSD, with 
    \(\partial_\parallel\) the spatial derivative along the line-of-sight direction, \(\hat{\bm r}\), and \(V\) the peculiar velocity potential. 
    \item The third term is the lensing magnification contribution, sourced by the integrated matter density along the line of sight, i.e.\ the weak lensing convergence \(\kappa\), modulated by the so-called magnification bias, \(\bq\), which respectively take the forms
    \begin{align}
        \kappa(\bm r)&=\int_0^r\de x\;\left(r-x\right)\,\frac{x}{r}\,\nabla^2_\perp\Upsilon(\hat{\bm r},r=x)\;,\label{eq:kappa}\\
        \bq(z)&=2\left[1-\left.\frac{\partial\ln\bar n_{\rm g}(z;F>F_{\rm cut})}{\partial\ln F}\right|_{F_{\rm cut}}\right]\;,\label{eq:bmag}
    \end{align}
    with \( r(z)\) the radial comoving distance to redshift \(z\), such that \(\de r=\de z/H(z)\) and \(H(z)=(1+z)\cH(z)\), \(\nabla^2_\perp\) the Laplacian on the transverse screen space, $\Upsilon=(\Phi+\Psi)/2$ the Weyl potential, where $\Phi$ and $\Psi$ are the two Bardeen potentials of the perturbed metric, and \(\bar n_{\rm g}\) the mean redshift-space comoving number density of galaxies, which is a function of redshift and flux \(F\) (equivalently luminosity, or magnitude). Here, $F_{\rm cut}$ represents the flux value that a galaxy should have in order to be detected by the adopted instrument.
    \item The penultimate term in \autoref{eq:Delta_g} gathers all the local contributions at the source, such as Sachs-Wolfe and Doppler terms, and reads
    \begin{equation}
        \Delta_{\rm loc}=(3-\be)\cH\,V+A\,\partial_\parallel V-\bq\Phi+(1
        -A)\Psi+\frac{\Phi^\prime}{\cH}\;,\label{eq:Delta_loc}
    \end{equation}
    with
    \begin{align}
        \be(z)&=-\frac{\partial\ln\bar n_{\rm g}(z)}{\partial\ln(1+z)}\label{eq:bevo}
    \end{align}
    usually referred to as the evolution bias,\textsuperscript{\ref{Footnote1}}
    \begin{equation}
        A\equiv\be
        +\bq-2
        -\frac{\cH^\prime}{\cH^2}
        -\frac{\bq}{\cH r}\;,
    \end{equation}
    and a prime denoting derivation with respect to conformal time.
    \item The last term, on the other hand, collects all non-local contributions, such as time delay and integrated Sachs-Wolfe type terms, and reads
    \begin{equation}
        \Delta_{\rm int}=
        2\frac{\bq}{r}\int_0^r\de x\;\Upsilon-2A\int_0^r\de x\;\Upsilon^\prime\;.
    \end{equation}
\end{enumerate}

\subsection{The exact expression}\label{sec:std}

The exact linear harmonic-space angular power spectrum of the observed galaxy number count fluctuations between two (infinitesimally thin) redshift slices at \(z\) and \(z^\prime\), \(\cex(z,z^\prime)\), is then obtained by expanding \autoref{eq:Delta_g} in spherical harmonics, and taking the ensemble average
\begin{equation}
    \left\langle\Delta_{{\rm g},\ell m}(z)\Delta^*_{{\rm g},\ell^\prime m^\prime}(z^\prime)\right\rangle\equiv\delta^{\rm K}_{\ell\ell^\prime}\delta^{\rm K}_{mm^\prime}\cex(z,z^\prime),
\end{equation}
with $\delta^{\rm K}$ the Kronecker delta symbol. This leads to the expression (`Ex' meaning `exact')
\begin{equation}
    \cex(z,z^\prime)=4\pi\int\de\ln k\;\cW_\ell^{\rm g}(k;z)\,\cW_\ell^{\rm g}(k;z^\prime)\,\cP_\zeta(k)\;,\label{eq:Cl_exact}
\end{equation}
with \(\cW_\ell^{\rm g}\) the kernel of galaxy clustering, encompassing contributions from all terms present in \autoref{eq:Delta_g}, and \(\cP_\zeta(k)= A_{\rm s}\,k^{n_{\rm s}-1}\) the power spectrum of primordial curvature perturbations, \(A_{\rm s}\) and \(n_{\rm s}\) respectively being its amplitude and spectral index.

For a full expression for \(\cW_\ell^{\rm g}\), we can write
\begin{equation}
    \cW_\ell^{\rm g}=\cW_\ell^{\rm g,den}+\cW_\ell^{\rm g,vel}+\cW_\ell^{\rm g,len}+\cW_\ell^{\rm g,rel}\;,\label{eq:W_full}
\end{equation}
with \(\cW_\ell^{\rm g,vel}=\cW_\ell^{\rm g,RSD}+\cW_\ell^{\rm g,Dop}\) the term related to galaxies' velocities, where \citep[see e.g.][]{DiDio:2013bqa}
\begin{equation}
    \cW_\ell^{\rm g,den}(k;z)=\bg(k,z)\,T_\delta(k,z)\,j_\ell\left[k r(z)\right]\;,\label{eq:W_den}
\end{equation}
\begin{equation}
    \cW_\ell^{\rm g,RSD}(k;z)=\frac{k}{\cH(z)}\,T_V(k,z)\,j^{\prime\prime}_\ell\left[k r(z)\right]\;,\label{eq:W_RSD}
\end{equation}    
\begin{align}
    \cW_\ell^{\rm g,Dop}(k;z)&=\bigg\{\left[\be(z)-3\right]\,\frac{\cH(z)}{k}\,j_\ell\left[k r(z)\right]\nonumber\\
    &\phantom{=\bigg\{}-A(z)\,j^{\prime}_\ell\left[k r(z)\right]\bigg\}\,T_V(k,z)\;,\label{eq:W_Dop}
\end{align}
\begin{multline}
    \cW_\ell^{\rm g,len}(k;z)=\ell\,(\ell+1)\,\bq(z)\\
    \times\int_0^{ r(z)}\de x\;\frac{ r(z)-x}{r(z)x}\,T_\Upsilon(k,r=x)\,j_\ell(kx)\;,\label{eq:W_len}
\end{multline}
\begin{align}
    \cW_\ell^{\rm g,rel}(k;z)&=\Big\{\left[1-A(z)\right]\,T_\Psi(k, z)-2\,\bq(z)\,T_\Phi(k, z)\nonumber\\
    &\phantom{=\big\{}+\frac1{\cH(z)}\,T_{\Phi^\prime}(k,z)\Big\}j_\ell\left[k r(z)\right]\nonumber\\
    &\phantom{=}+2\,\frac{\bq(z)}{ r(z)}\int_0^{ r(z)}\de x\;T_\Upsilon(k,r=x)j_\ell(kx)\nonumber\\
    &\phantom{=}-2\,A(z)\int_0^{ r(z)}\de x\;T_{\Upsilon^\prime}(k,r=x)j_\ell(kx)\;.\label{eq:W_rel}
\end{align}
In the equations above, $T_X$ denotes the transfer function describing the evolution of the random variable $X$ and $T_X(k,z)\equiv T_X[k,r(z)]$. Note that, with a slight abuse of notation, a prime applied to a spherical Bessel function denotes a derivative with respect to its argument.

In harmonic-space analyses, it is customary to subdivide the observed source population into redshift bins. This is done, for instance, to reduce the dimensionality of the data vector---and consequently the covariance matrix---with the aim of reducing in turn the computational complexity of the problem. Otherwise, redshift information for the observed galaxies might be too poor to allow us to pin them down in the radial direction, as is the case with photometric redshift estimation. In this case, galaxies are usually binned into $\mathcal O(1)-\mathcal O(10)$ bins spanning the observed redshift range. Whatever the reason, in practice this corresponds to having
\begin{equation}
    \cijex=4\pi\int\de\ln k\;\cW_{i\ell}^{\rm g}(k)\,\cW_{j\ell}^{\rm g}(k)\,\cP_\zeta(k)\;,\label{eq:Cijl_exact}
\end{equation}
where
\begin{equation}
    \cW_{i\ell}^{\rm g}(k)=\int\de z\;\cW_\ell^{\rm g}(k;z)\,n_i(z)\;,
\end{equation}
with $n_i(z)$ the galaxy redshift distribution in the $i$th redshift bin, normalised to unit area.

\subsection{Widely used approximations}\label{sec:approx}

The computation of harmonic-space power spectra has to be performed following the triple integral of \autoref{eq:Cijl_exact} and the equations giving the kernel $\cW_{i\ell}^{\rm g}$. However, such  an integration is numerically cumbersome, especially because of the presence of spherical Bessel functions---highly oscillatory functions whose amplitude and period vary significantly with the argument of the function. As a consequence, numerical integration has to be performed with highly adaptive methods, at the cost of computation speed. 
Over the years, various algorithms have been proposed with the aim of speeding up the computation of harmonic-space power spectra. Mostly, they rely on FFT/FFTLog methods \citep[see e.g.][]{2017JCAP...11..054A,2017A&A...602A..72C,2018PhRvD..97b3504G}.

On the other hand, the full computation is not always necessary, and approximations can be made to speed up the numerical evaluation, e.g.\ by applying the Limber or the flat-sky approximations \citep[often erroneously thought to be the same, see e.g.][]{Matthewson:2020rdt}. Here, we shall focus on the former, which is by far the most widely employed. It relies on the following property of spherical Bessel functions,
\begin{equation}
    j_\ell(x)\underset{\ell\gg1}{\longrightarrow}\sqrt{\frac{\upi}{2\ell+1}}\delta_{\rm D}\left(\ell+\frac{1}{2}-x\right)\;,\label{eq:Limber}
\end{equation}
where $\delta_{\rm D}$ is a Dirac delta.\footnote{Note that the $+1/2$ term comes from the relation between a spherical Bessel function of order $\ell$, $j_\ell$, and the ordinary Bessel function of order $L=\ell+1/2$, $J_L$.} By performing the substitution of \autoref{eq:Limber} into \autoref{eq:Cijl_exact}, which contains $j_\ell$ through the $\cW_{i\ell}^{\rm g}(k)$, we can effectively get rid of two integrations, thus boosting significantly the speed of the computation.

Moreover, the relative importance of the different terms in \autoref{eq:W_full} depends on various, survey-dependent factors. For instance, RSD are mostly washed out for broad redshift bins, whereas, on the contrary, lensing magnification favours them. Similarly, the Doppler contribution decays quickly as the redshift of the shell grows, whilst integrated terms like lensing gain in weight. Lastly, the importance of the various effects also varies with the scales of interest, as can be seen by the $\cH/k$ factors in \autoref{eq:W_den} to \autoref{eq:W_rel}. Moreover, note that at first order in cosmological perturbation theory, the Einstein equations fix $V\sim\delta/k$ and $\Phi\sim\Psi\sim\delta/k^2$. All combined, this makes $\cW^{\rm g,rel}_\ell$ important only on very large scales.

For these reasons, galaxy clustering in harmonic space is customarily restricted to Newtonian density fluctuations alone, leading to the well-known expression for the approximated angular spectra (`Ap' standing for `approximated')
\begin{equation}
    \cijapp=\int\de z\;\frac{\left[H(z)\,\bg(z)\right]^2\,n^i(z)\,n^j(z)}{r^2(z)}\,P_{\rm lin}\!\left[\frac{\ell+1/2}{r(z)},z\right]\;,\label{eq:Cijl_approx}
\end{equation}
where $P_{\rm lin}(k,z)$ is the linear matter power spectrum, and for now we have assumed that linear galaxy bias is only redshift-dependent. Let us emphasise that this approximation, and in particular the neglection of RSD, is ofttimes common in harmonic-space analyses of galaxy clustering \citep[see e.g.][]{2012MNRAS.421..251G,2018MNRAS.476.4662V,2021arXiv210513549D}, albeit with noticeable exceptions \citep[][]{2007MNRAS.378..852P,2019MNRAS.485..326L,2021A&A...646A.129J,2021arXiv210700026T}. Oppositely, real- and Fourier-space analyses do customarily account for RSD.

The actual accuracy of such an approximation, however, cannot be estimated a priori, since it strongly depends on the integrand of \autoref{eq:Cijl_approx}. In particular, \autoref{eq:Cijl_approx} is known to agree well with the exact expression of \autoref{eq:Cijl_exact} if the kernel of the integral is broad in redshift. Moreover, the Limber approximation works better at low redshift than at high redshift, because the higher the redshift, the larger the scale subtended by a given angular separation; in other words, the minimum multipole for which the Limber approximation agrees well with the exact solution increases with redshift.

\begin{figure}
    \centering
    \includegraphics[width=0.48\textwidth]{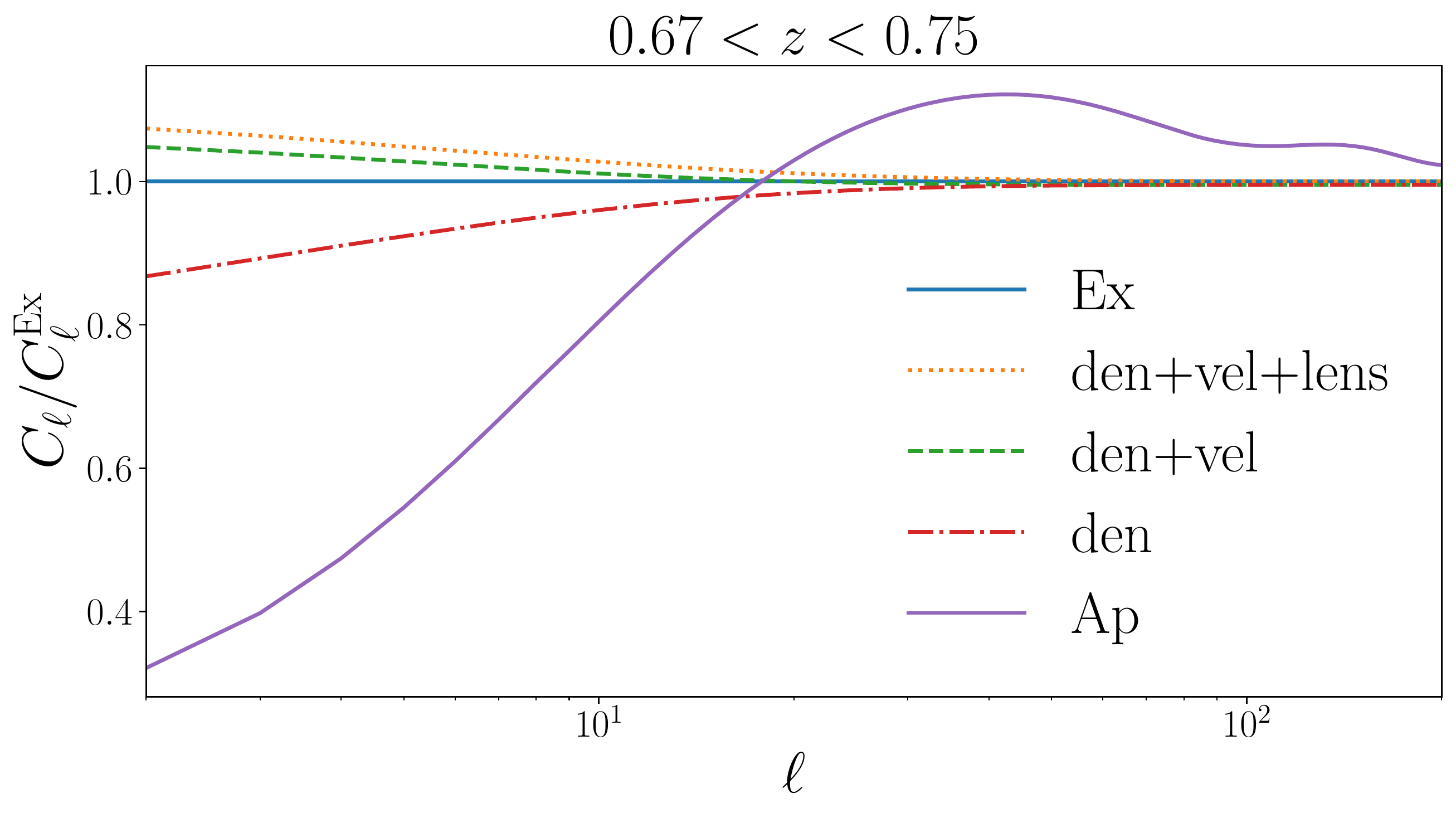}
    \caption{Ratio of the approximated $C_\ell$ to the exact $\cex$ of \autoref{eq:Cijl_exact}. The labels of the different curves correspond to the contributions that enter the window function in \autoref{eq:W_full}. None of the spectra considered here use the Limber approximation, except the `Ap' spectrum, which corresponds to the fully approximated $\capp$ of \autoref{eq:Cijl_approx}. The spectra shown here refer to the auto-correlation in a redshift bin with $0.67<z<0.75$, using the survey specifications discussed in \autoref{sec:survey}.}
    \label{fig:contributions}
\end{figure}

In \autoref{fig:contributions}, we highlight the contributions of the different terms to the final spectra, by showing the ratio of approximated spectra to the exact ones. We consider here the auto-correlation spectra in a redshift bin with $0.67<z<0.75$, using the survey specifications we later discuss in \autoref{sec:survey}. None of the spectra shown in the figure use the Limber approximation, except the `Ap' spectrum, which corresponds to \autoref{eq:Cijl_approx}. We notice how removing different terms makes the theoretical prediction move away from the exact one, although only at very large scales and not in a dramatic way, even when only the density term of \autoref{eq:W_den} is kept. However, once the Limber approximation is used, the predictions significantly depart form the exact spectrum over a wide range of multipoles.

\section{Survey specifications}\label{sec:survey}
In the coming decade, several planned surveys of the cosmic large-scale structure will provide us with observations of the galaxy distribution with unprecedented sensitivity at very large scales. It is therefore crucial to assess how the common approximations described in \autoref{sec:GC} will impact the accuracy of the results we will be able to obtain. Therefore, in this paper we adopt the specifications of a very deep and wide galaxy clustering survey with high redshift accuracy. We emphasise that we are not interested in forecasts for a specific experiment, but rather in assessing whether and how much various approximations affect the final science output. For this reason, we shall focus on an idealised survey, loosely inspired by the envisaged future construction phase of the SKAO. Specifically, we consider an HI-galaxy redshift survey, assuming that the instrument will be able to provide us with spectroscopic measurements of the galaxies' redshifts through the detection of the HI emission line in the galaxy spectra. Therefore, for the purposes of the harmonic-space tomographic studies we focus on in this paper, we shall consider the error on such redshift measurements to be negligible.

Here, we follow the prescription and fitting functions of \citet{2015MNRAS.450.2251Y} to characterise the source galaxy distribution as a function of both redshift and flux limit. The latter will be particularly important in determining the magnification bias of the sample. Calculations in \citet{2015MNRAS.450.2251Y} were based on the $S^3$-SAX simulations by \citet{Obreschkow:2009ha} and assumed that any galaxy with an integrated line flux above a given signal-to-noise ratio threshold would be detected. The fitting formulae adopted here are
\begin{align} 
\frac{\de N_{\rm gal}}{\de z} &= 10^{c_1}z^{c_2} \exp({-c_3 z})\,\mathrm{deg^{-2}}\;,\label{eq:hi_dndz}\\
\bg(z) &= c_4 \exp(c_5\,z)\;,\label{eq:linbias}
\end{align}
where $N_{\rm gal}$ is the total number of galaxies in the entire redshift range of the survey, and parameters \(c_i\) can be found in \citet[][]{2015MNRAS.450.2251Y} for a wide range of flux thresholds, from \(0\) to \(200\,\mathrm{\mu Jy}\). We show in \autoref{tab:survey} values of \(c_i\) used in the present work, corresponding to those used in \citet{Sprenger:2018tdb} and obtained in \citet{Bull:2015lja} as a result of fitting these functions to the expected galaxy number density given the survey design.

\begin{figure}
    \centering
    \includegraphics[width=0.48\textwidth]{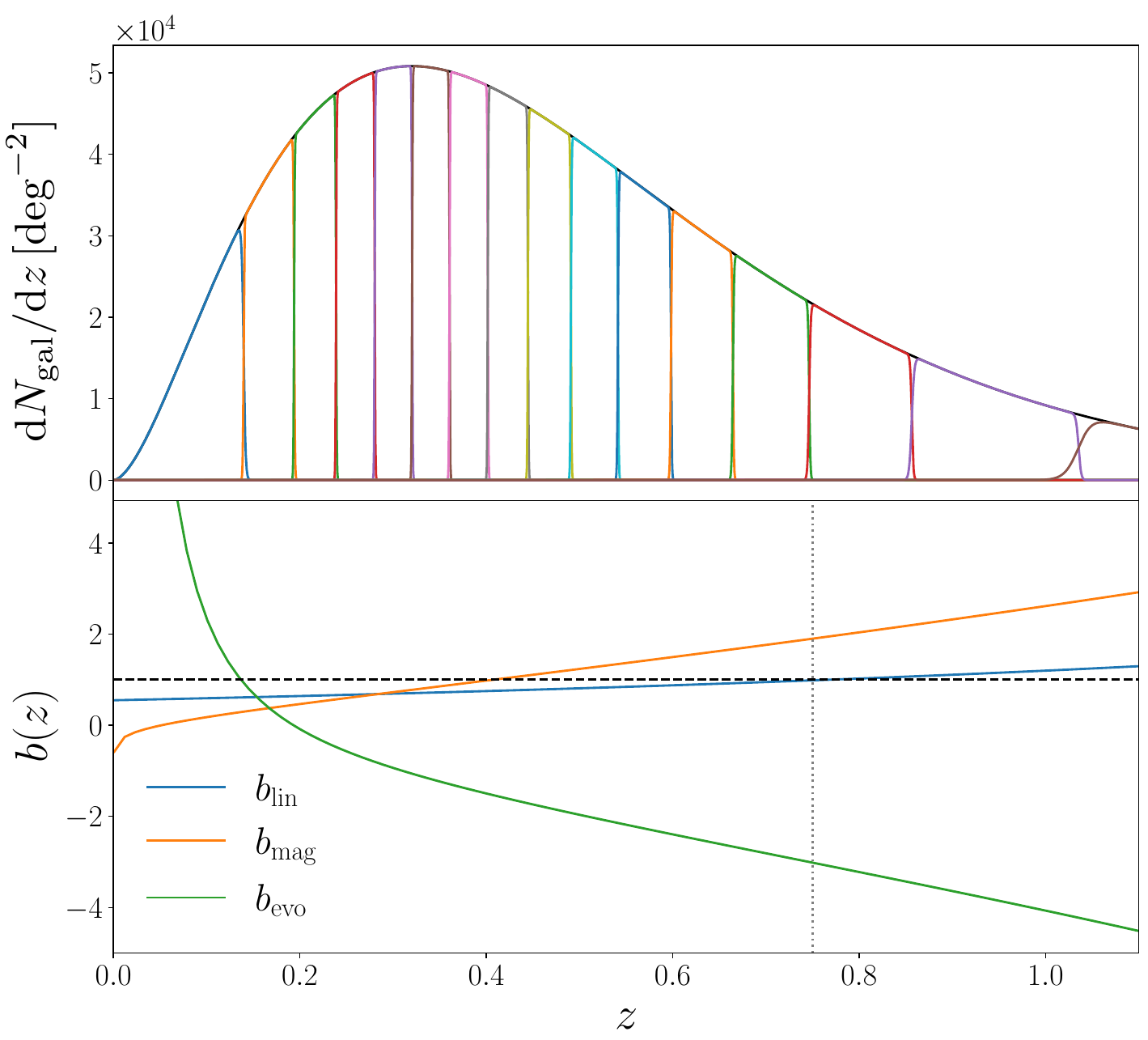}\\
    \caption{{\it Upper panel:} Galaxy distribution as described by \autoref{eq:hi_dndz} (in black) with the limits of the equipopulated redshift bins considered in the present paper (in colour). {\it Lower panel:} Trends in redshift for the linear galaxy bias of \autoref{eq:linbias} (blue curve), the magnification bias of \autoref{eq:bmag} (orange curve) and the evolution bias of \autoref{eq:bevo} (green curve). The intersection between the horizontal dashed and vertical dotted black lines shows where the linear galaxy bias crosses unity.}
    \label{fig:specs}
\end{figure}

Given the galaxy distribution of \autoref{eq:hi_dndz}, we focus on the redshift range $0.001<z<1.1$ with $N_{\rm gal}$ given in \autoref{tab:survey}, and divide it into $N_{\rm bin}=15$ redshift bins assuming that each one contains the same number of galaxies (see the upper panel of \autoref{fig:specs}). In the lower panel of \autoref{fig:specs}, we show the redshift evolution of the linear galaxy bias, the magnification bias and the evolution bias given, respectively, by \autoref{eq:linbias}, \autoref{eq:bmag} and \autoref{eq:bevo}, for the survey under consideration.

Using these survey specifications, we create a simulated data set for galaxy clustering observations; we calculate the exact angular power spectra $\cex$, described in \autoref{sec:GC}, in a fiducial cosmology and we add to these the noise computed using the survey specifications. For the rest of this paper we use the calculations of the exact and approximated power spectra as implemented in \texttt{CAMB}\footnote{\url{https://github.com/cmbant/CAMB}} \citep{Lewis:1999bs,Howlett:2012mh}. We assume a $\Lambda$CDM cosmology with fiducial values of parameters given in \autoref{tab:survey}, where $\omega_{\rm b}$ and $\omega_{\rm c}$ are the baryon and cold dark matter physical energy densities, respectively, $h$ is the reduced present-day Hubble expansion rate, $A_{\rm s}$ and $n_{\rm s}$ are, respectively, the amplitude and spectral index of the primordial curvature power spectrum, and $\sum{m_\nu}$ is the sum of the neutrino masses.

\begin{table}
    \LARGE
    \centering
    \resizebox{\columnwidth}{!}{
    \begin{tabular}{|c|c|c|c|c|c|c|c|c|}
    \hline
    \multicolumn{9}{c}{Survey specifications}\\
    \hline
        $N_{\rm gal}$ & $f_{\rm sky}$ & $z_{\rm min}$ & $z_{\rm max}$ & $c_1$  & $c_2$  & $c_3$  & $c_4$  & $c_5$\\
        $9.4\times10^8$        & $0.7$         & $0.001$       & $1.1$         & $6.32$ & $1.74$ & $5.42$ & $0.55$ & $0.78$\\
    \hline
    \hline
    \multicolumn{9}{c}{Fiducial cosmology}\\
    \hline
         $\omega_{\rm b}$ & $\omega_{\rm c}$ & $h$ & $A_{\rm s}\times 10^9$ & $n_{\rm s}$  & $\sum{m_\nu}$ [eV]    & $w$ & $f_{\rm NL}$ & \\
         $0.02245$        & $0.12056$         & $0.67$       & $2.126$         & $0.96$ & $0.06$  & $-1$ & $0$ & \\
    \hline    
    \end{tabular}}
    \caption{Survey specifications and fiducial cosmology used in the present work to obtain the mock data set and experimental noise.}
    \label{tab:survey}
\end{table}

\section{Case studies}\label{sec:cases} 

We study four representative cosmological models in order to demonstrate how the approximations of \autoref{sec:approx} can bias the estimation of cosmological parameters using a next-generation survey able to access ultra-large scales, as described in \autoref{sec:survey}, and how the method we present in this paper debiases the constraints while keeping the computational cost of the parameter estimation procedure significantly lower than that of an exact analysis. These four models are the standard $\Lambda$CDM model and three of its minimal extensions, where either the dark energy equation of state $w$ or the sum of the neutrino masses $\sum{m_\nu}$ or the local primordial non-Gaussianity (PNG) parameter $f_\mathrm{NL}$ is allowed to vary as an additional free parameter. We denote these extensions by $w$CDM, $\Lambda$CDM+$m_\nu$ and $\Lambda$CDM+$f_\mathrm{NL}$, respectively.

\subsection{Standard model and its simple extensions}\label{sec:lcdm}

We specify the standard $\Lambda$CDM model by the five free parameters $\omega_{\rm b}$, $\omega_{\rm c}$, $h$, $A_{\rm s}$ and $n_{\rm s}$.\footnote{Note that $\Lambda$CDM also requires the reionization optical depth $\tau$ as a free parameter. However, we do not vary $\tau$ in our analysis as we do not expect the large-scale observables to constrain this quantity.} Following \citet{2020A&A...641A...6P}, we fix the value of $\sum{m_\nu}$ to $0.06~\mathrm{eV}$ for $\Lambda$CDM. The parameters $\{\omega_{\rm b},\omega_{\rm c},h,A_{\rm s},n_{\rm s}\}$ affect the angular power spectra of the observed galaxy number count fluctuations differently and on different angular scales. Here we are interested in ultra-large scales, which are expected to be particularly sensitive to the parameters that quantify cosmic initial conditions, i.e.\ $A_{\rm s}$ and $n_{\rm s}$. 

\begin{figure*}
    \centering
    \includegraphics[width=0.45\textwidth]{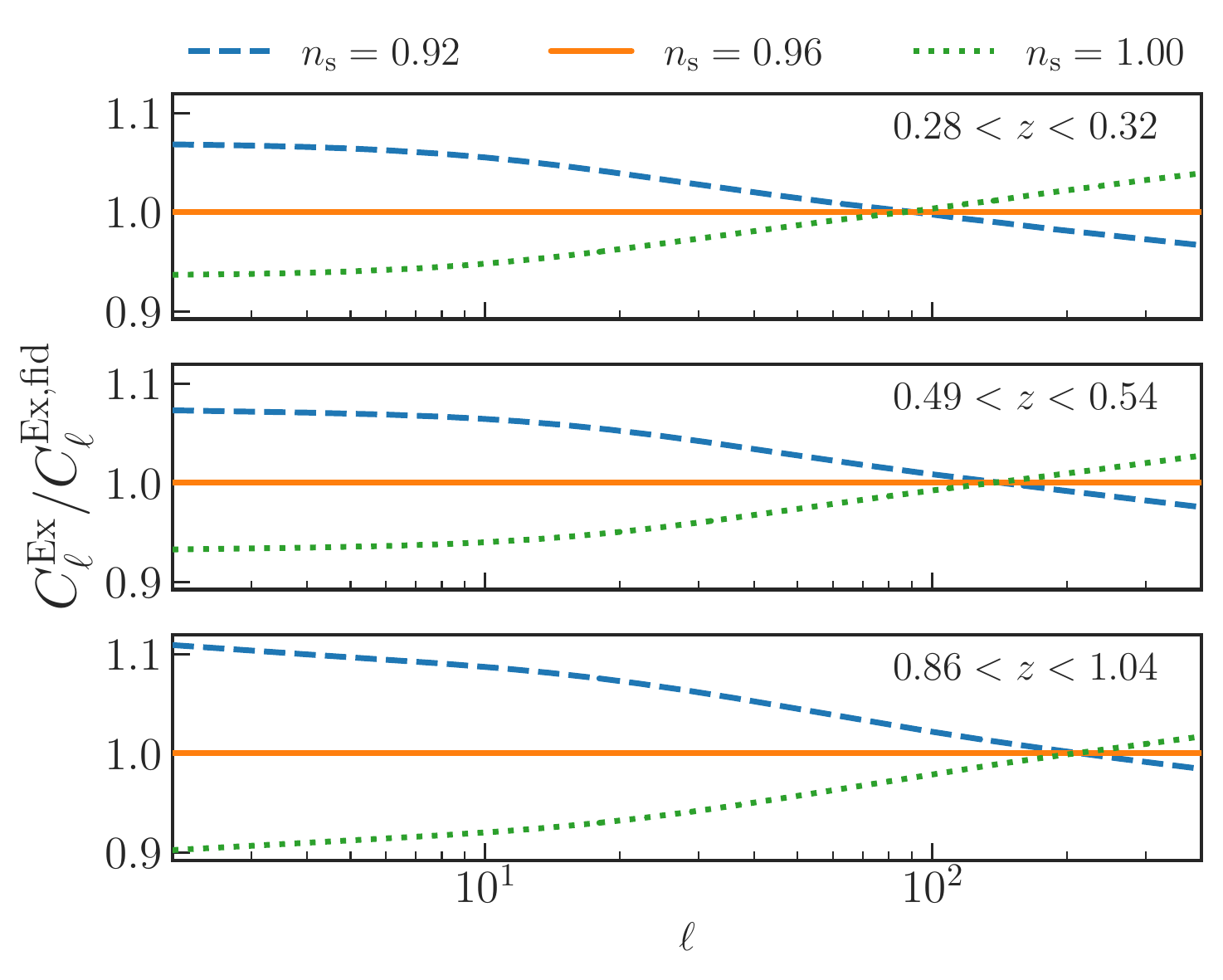}
    \includegraphics[width=0.45\textwidth]{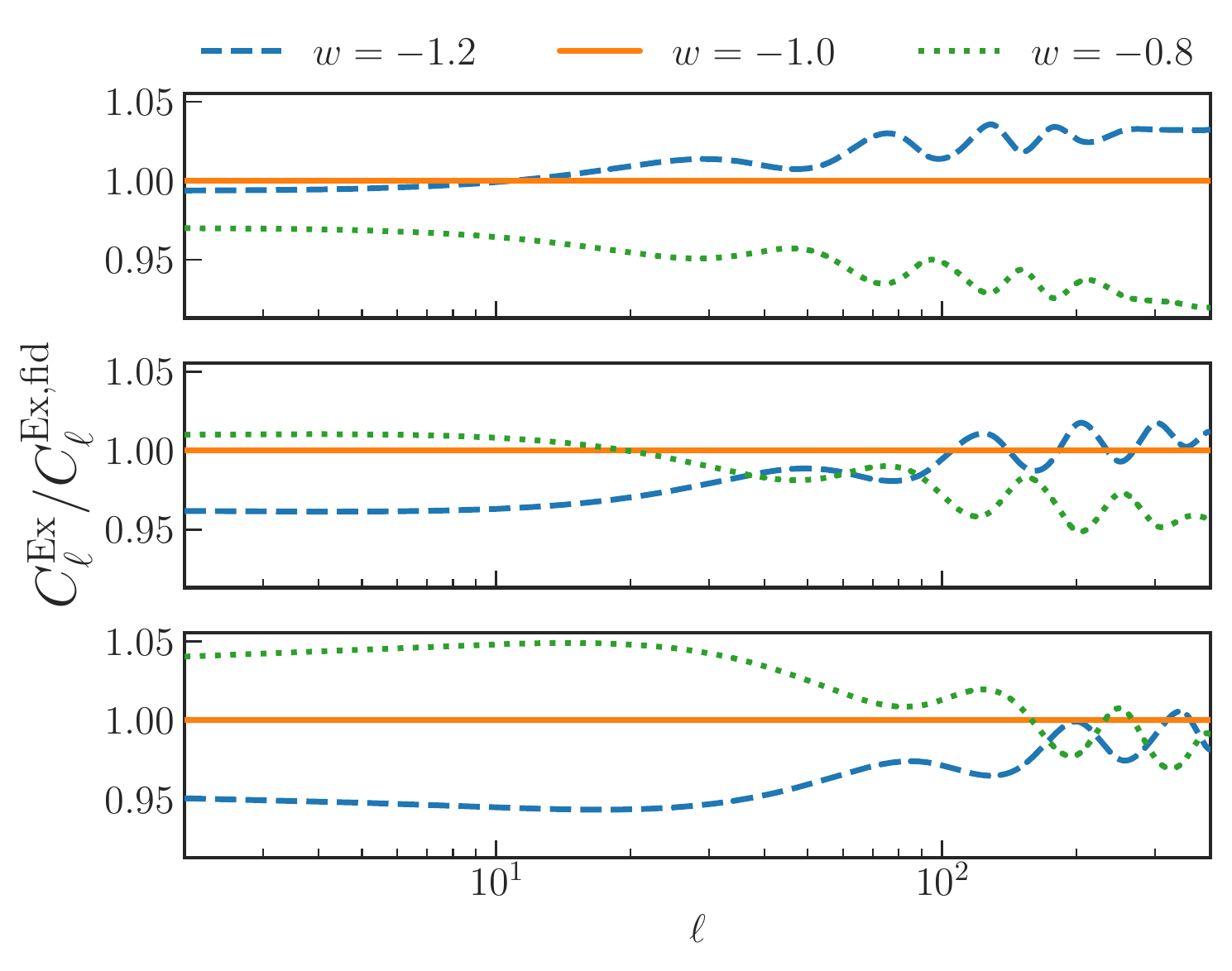}
    \includegraphics[width=0.45\textwidth]{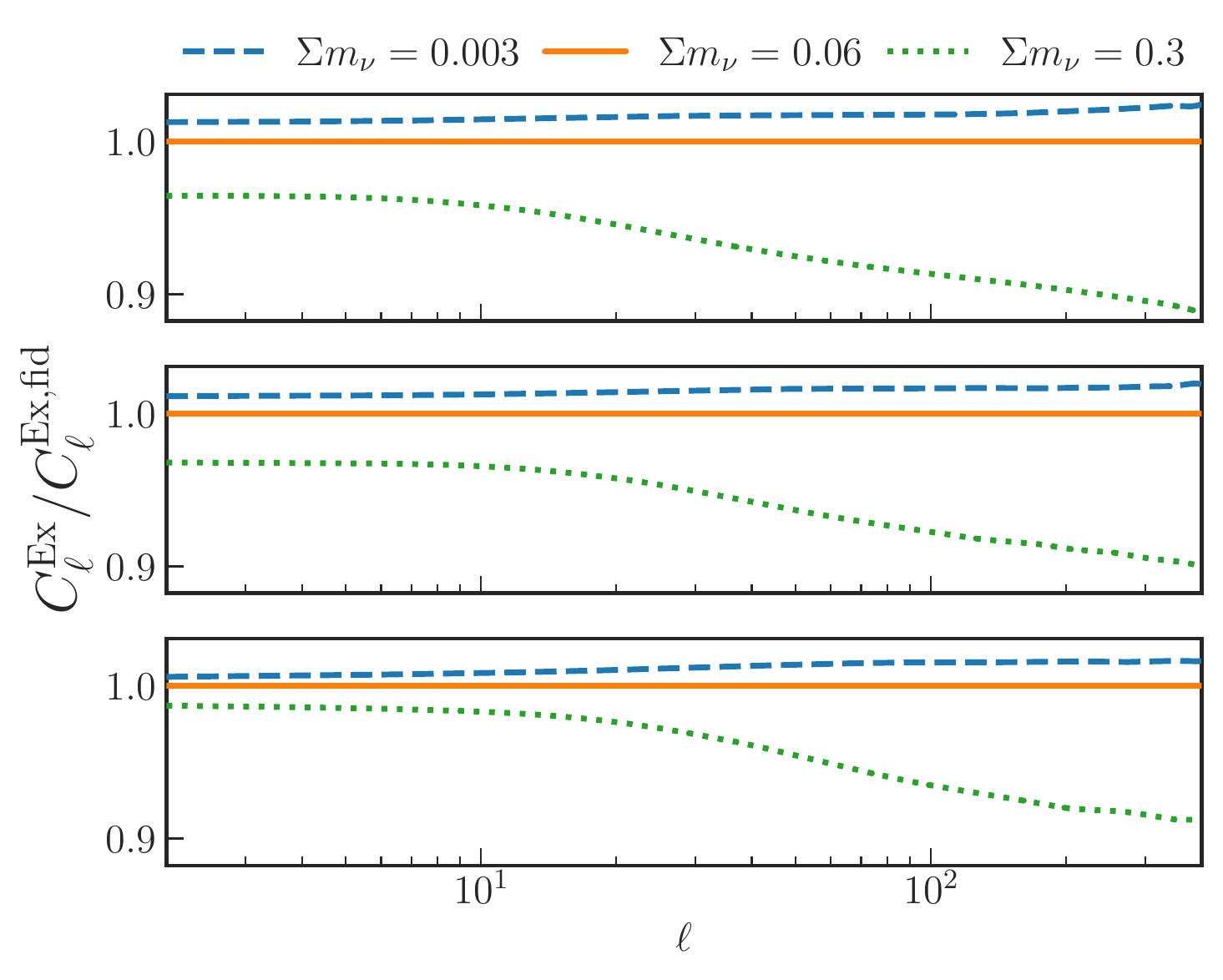}
    \includegraphics[width=0.45\textwidth]{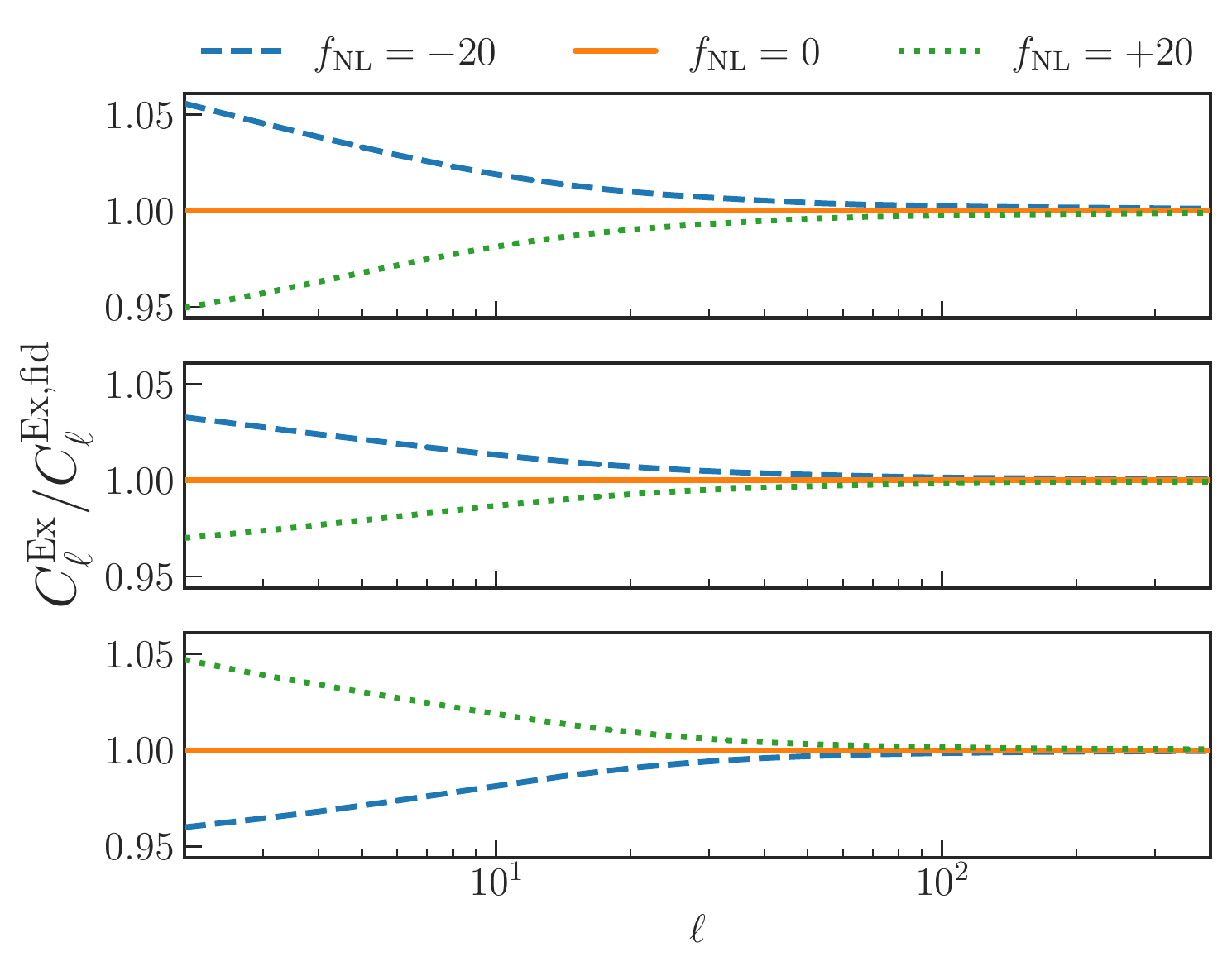}
    \caption{Effects of cosmological parameters on the angular power spectrum of observed galaxy number count fluctuations, $C_\ell$, on large scales. The four panels depict the effects of: the primordial scalar spectral index $n_{\rm s}$ in $\Lambda$CDM (upper left panel); the dark energy equation of state $w$ in $w$CDM (upper right panel); the sum of the neutrino masses $\sum{m_\nu}$ in $\Lambda$CDM+$m_\nu$, with values in $\mathrm{eV}$ (lower left panel); and the local primordial non-Gaussianity $f_\mathrm{NL}$ in $\Lambda$CDM+$f_\mathrm{NL}$ (lower right panel). All the power spectra are exact, i.e.\ no approximations are made in their computations, and they are shown in comparison with the fiducial $\Lambda$CDM spectra with $n_{\rm s}=0.96$, $w=-1$, $\sum{m_\nu}=0.06~\mathrm{eV}$ and $f_\mathrm{NL}=0$. Each panel contains three sets of spectra computed for the three redshift bins $5$, $10$ and $15$, corresponding to low, medium and high redshifts (from top to bottom in each panel). The redshift range of each bin is indicated in the respective plot in the upper left panel.}
    \label{fig:theory}
\end{figure*}

In order to illustrate the large-scale effects of the parameters, we show, as an example, in the upper left panel of \autoref{fig:theory} the impact of varying the scalar spectral index $n_{\rm s}$ on the power spectrum at angular scales larger than $\ell=400$ computed at redshift bins $5$, $10$ and $15$ (corresponding to redshift ranges $0.28<z<0.32$, $0.49<z<0.54$ and $0.86<z<1.04$, respectively) as given in the upper panel of \autoref{fig:specs}. For each redshift bin, the corresponding galaxy number count power spectra for three values of $n_{\rm s}=0.92$, $n_{\rm s}=0.96$ and $n_{\rm s}=1$ are shown relative to the spectrum for $n_{\rm s}=0.96$, which we use as our fiducial value in the rest of this paper. Note that these spectra are all exact, i.e.\ they are computed without making any approximations. As the figure shows, in all the redshift bins, the lower the value of $n_{\rm s}$, the more enhanced the power spectra at ultra-large scales, namely scales with $\ell\lesssim \mathcal{O}(100)$, while we see the opposite effect at smaller scales. This is because the smaller the value of $n_{\rm s}$, the steeper (or more `red-tilted') the primordial power spectrum, resulting in larger amplitudes of fluctuations at extremely large scales. This steeper spectrum will then lead to suppression of amplitudes at scales smaller than some `pivot’ scale. Note, however, that the enhancement/suppression on large scales is not physical, as it depends on the scale used as a pivot---namely, fixing either $A_{\rm s}$ or $\sigma_8$ (amplitude of the linear power spectrum on the scale of $8 \,h^{-1}\,\mathrm{Mpc}$) as a fundamental parameter.

The figure for $n_{\rm s}$ already shows the importance of correctly computing the angular power spectra for accurately estimating the cosmological parameters using ultra-large-scale information. As the figure shows, even changing $n_{\rm s}$ to the extreme values of $0.92$ and $1$, both of which having aleady been ruled out by the current constraint $n_{\rm s}\approx0.965\pm0.0042$ \citep{2020A&A...641A...6P}, changes the power spectra by $\mathcal{O}(10\%)$. On the other hand, as we will see in \autoref{sec:results}, the approximations of \autoref{sec:approx} may easily result in $>\mathcal{O}(10\%)$ errors in the computation of the spectra on large scales, which will then lead to inaccurate, or biased, estimates of parameters like $n_{\rm s}$.

An inaccurate estimation of a cosmological parameter can also result in a false detection of new physics when there is none, or in no detection when there is. In order to demonstrate this problem, we present in the upper right and lower left panels of \autoref{fig:theory} the effects of the two important non-standard cosmological parameters, $w$ and $\sum{m_\nu}$, on the power spectrum at large scales for $w$CDM and $\Lambda$CDM+$m_\nu$, the two simple extensions of $\Lambda$CDM that we introduced earlier. The panels again depict the spectra for the three redshift bins $5$, $10$ and $15$, with the additional parameters $w$ and $\sum{m_\nu}$ of the two extensions set to $\{-1.2,-1,-0.8\}$ and $\{0.003,0.06,0.3\}$, respectively. Note that throughout this paper, we always use $w=-1$ and $\sum{m_\nu}=0.06$ as the fiducial values for these parameters.

We notice that changing the value of $w$ has a few large-scale effects. First of all, setting $w$ to a value smaller or larger than $-1$ does not affect the spectra similarly in different redshift bins. Focusing first on the $w=-1.2$ case, which corresponds to a phantom dark energy, we see that the spectra are all suppressed at ultra-large scales compared to the standard $w=-1$ case, and by increasing the bin's redshift, not only does the range of the suppressed power extend to smaller scales, but also the higher the redshift, the more suppressed the spectra (on all scales). The effect is the opposite for the $w=-0.8$ case, and increasing the bin's redshift results in more enhanced spectra compared to the baseline $w=-1$. The $w\neq-1$ enhancement/suppression of power and its redshift dependence can be explained for smaller scales by the fact that the linear growth rate of the large-scale structure, $f$, is significantly affected by $w$, especially at low redshifts, where dark energy becomes more important \citep[see e.g.][]{2010deto.book.....A}. At any given redshift $z$, a larger $w$ makes the dark energy component more important compared to the matter component, and since the growth rate $f(z)$ increases by increasing the dark matter component, it decreases by increasing $w$. This is exactly what we see in \autoref{fig:theory} for the three values of $w=-1.2$, $w=-1$ and $w=-0.8$. We also see that, as expected, the differences between the three spectra at smaller scales are significantly reduced when we increase the bin's redshift. The dependence of the power spectrum on the value of $w$ is, however, much more involved for very large scales, as the spectrum on those scales is determined by a combination of different $w$-dependent effects, such as the integrated Sachs Wolfe effect. Finally, in all the three bins of the upper right panel of \autoref{fig:theory}, the oscillatory features in the ratios $C_\ell/C_\ell^\mathrm{fid}$ are a consequence of the fact that the baryon acoustic oscillations shift towards smaller scales with increasing redshift for both $w=-1.2$ and $w=-0.8$.

When considering the sum of the neutrino masses, we see that increasing $\sum{m_\nu}$ results in the suppression of power on all scales and in all redshift bins, although this suppression is significantly stronger at smaller scales (or higher multipoles). There are several reasons for the small-scale reduction of the power spectra in the presence of massive neutrinos \citep[see e.g.][]{Lesgourgues:2012uu}, the most important of which is the absence of neutrino perturbations in the total power spectrum and a slower growth rate of matter perturbations at late times. On extremely large scales, however, neutrino free-streaming can be ignored \citep[see e.g.][]{Lesgourgues:2012uu} and neutrino perturbations are therefore indistinguishable from matter perturbations. The power spectra then depend only on the total matter+neutrino density fraction today and on the primordial power spectrum. The small suppression of the angular power spectra at ultra-large scales, as seen in \autoref{fig:theory}, is therefore because of the contribution of massive neutrinos to the total density parameter $\Omega_\mathrm{m}$.

\subsection{Primordial non-Gaussianity and scale-dependent bias}\label{sec:fnl}
An important extension of the standard $\Lambda$CDM model for our studies of ultra-large scales is $\Lambda$CDM+$f_\mathrm{NL}$, where the parameter $f_\mathrm{NL}$ is added to the model in order to capture the effects of a non-zero local primordial non-Gaussianity. It has been shown \citep{Dalal:2007cu,Matarrese:2008nc,Slosar:2008hx} that a local PNG modifies the Gaussian bias by contributing a scale-dependent piece of the form
\begin{equation}
\Delta b(z,k)=3[b_\mathrm{lin}(z)-1]\frac{\delta_{\rm c}\,\Omega_\mathrm{m}\,H_0^2}{k^2\,T(k)\,D(z)}f_\mathrm{NL}\,,\label{eq:scaledepbias}
\end{equation}
where $\Omega_\mathrm{m}$ is the present-day matter density parameter, $H_0$ is the value of the Hubble expansion rate today, $T(k)$ is the matter transfer function (with $T\to 1$ as $k\to 0$), $D(z)$ is the linear growth factor normalised to $(1+z)^{-1}$ in the matter-dominated Universe, and $\delta_{\rm c}\sim1.68$ is the (linear) critical matter density threshold for spherical collapse. The appearance of the $k^2$ factor in the denominator of \autoref{eq:scaledepbias} immediately tells us that ultra-large scales are the natural choice for placing constraints on $f_\mathrm{NL}$ using this scale-dependent bias, as the signal becomes stronger when $k\to 0$.

The lower right panel of \autoref{fig:theory} shows the effects of non-zero values of $f_\mathrm{NL}$ on the power spectrum at large scales---note that similar to the previous cases, the spectra are exact, i.e.\ no approximations are made in their computations. We first notice that, as expected, a non-zero $f_\mathrm{NL}$ only affects the ultra-large scales substantially, by enhancing or suppressing the power spectra, and that this happens in all the redshift bins shown in the figure. This again emphasises the importance of accurately and precisely measuring the power spectra at ultra-large scales, as even the unrealistically large values of $f_\mathrm{NL}=\pm20$ (see \citealt{2020A&A...641A...9P} for the current observational constraints on $f_\mathrm{NL}$) shown in the figure affect the spectra by only $<5\%$.

The figure also shows that a negative (positive) $f_\mathrm{NL}$ enhances (suppresses) the spectra for the two low-redshift bins $5$ and $10$, while the effect is the opposite for the high-redshift bin $15$. Here we explain the reason for this surprising but important feature. For that, let us investigate the redshift dependence of \autoref{eq:scaledepbias} for the full bias. The quantity $b_\mathrm{lin}$ is redshift-dependent and is given by \autoref{eq:linbias} for the survey we consider in this paper. As can be seen in the lower panel of \autoref{fig:specs}, the quantity $b_\mathrm{lin}-1$ is negative for $z\lesssim0.75$ and positive for $z\gtrsim0.75$, which means that a negative (positive) $f_\mathrm{NL}$ enhances the bias at low (high) redshifts and suppresses it at high (low) redshifts. Now looking at the upper panel of \autoref{fig:specs}, we see that the two upper bins of the lower right panel of \autoref{fig:theory} (bins 5 and 10) contain redshifts that are lower than $0.75$, while the lower bin (bin 15) includes redshifts higher than $0.75$.

It is, however, important to note that this is the case only if one assumes a $b_\mathrm{lin}-1$ factor in \autoref{eq:scaledepbias}, whose validity has been questioned in the literature (see e.g. \citealt{Barreira:2020ekm} and references therein). For this reason, we modify \autoref{eq:scaledepbias} as
\begin{equation}
\Delta b(z,k)=3f_\mathrm{NL}[b_\mathrm{lin}(z)-p]\delta_{\rm c}\frac{\Omega_\mathrm{m}H_0^2}{k^2T(k)D(z)}\,,\label{eq:scaledepbias-p}
\end{equation}
where $p$ is now a free parameter to be determined by cosmological simulations. It is argued by~\citet{Barreira:2020ekm} that $p=1$ for gravity-only dynamics and when universality of the halo mass function is assumed, while other values of $p$ provide better descriptions of observed galaxies where both of these assumptions are violated. Depending on the specific analysis and modelling, different values of $p$ have been obtained, e.g., \citet{Slosar:2008hx} and \cite{Pillepich:2008ka} showed that $p=1.6$ provides a better description of host halo mergers, while \citet{Barreira:2020kvh} showed that $p=0.55$ better describes 
IllustrisTNG-simulated stellar-mass-selected galaxies.

\section{Parameter estimation methodology}\label{sec:methods}
In this paper, we are interested in estimating the impact of large-scale effects and approximations on the estimation of cosmological parameters. In order to do so, we fit the mock data set obtained by the exact $\cex$ spectra as described in \autoref{sec:survey} using the $\capp$ spectra which make use of the several common approximations discussed in \autoref{sec:approx}. 

Throughout this work we rely on \texttt{CAMB} \citep{Lewis:1999bs,Howlett:2012mh} to compute the exact and approximated power spectra. We use a modified version of the code, following \citet{Camera:2014bwa}, when we consider the primordial non-Gaussianity parameter, $f_{\rm NL}$. We implement in the public code \texttt{Cobaya}\footnote{\url{https://github.com/CobayaSampler/cobaya}} \citep{Torrado:2020dgo} a new likelihood module which enables us to obtain from \texttt{CAMB} the approximated spectra $\capp$ and compare them with the mock data set. Such an analysis matches the approach commonly used for parameter estimation with galaxy number count data, where $\capp$ is computed at each step in the MCMC rather than $\cex$, as the computation of the latter is extremely time consuming and therefore unfeasible to repeat $\mathcal O(10^4)$ times.

For each point $\bm\Theta$ in the sampled parameter space, we compute the $\chi^2$ using the approach presented in \citet{2013JCAP...01..026A}, i.e.\
\begin{equation}\label{eq:like}
    \Delta\chi^2(\bm\Theta)=\sum_\ell{(2\ell+1)f_{\rm sky}\left(\frac{d_\ell^{\rm mix}(\bm\Theta)}{d_\ell^{\rm th}(\bm\Theta)}+\ln{\frac{d_\ell^{\rm th}(\bm\Theta)}{d_\ell^{\rm obs}}}-N_{\rm bin}\right)}\, ,
\end{equation}
where $N_{\rm bin}$ is the number of bins, and 
\begin{align}
d_\ell^{\rm th}(\bm\Theta) &= {\rm det}\left[\tilde{C}_{ij\ell}^{{\rm Ap}}(\bm\Theta)\right]\, , \\
d_\ell^{\rm obs} &= {\rm det}\left[\tilde{C}_{ij\ell}^{{\rm Ex}}(\bm\Theta^{\rm fid})\right]\,.
\end{align}
The tilde indicates that the used spectra contain an observational noise $N_{ij\ell}=\delta^{\rm K}_{ij}/n_i$, with $n_i$ the number of galaxies in the $i$th bin and $\delta^{\rm K}_{ij}$ the Kronecker delta, i.e.,\ $\tilde{C}_{ij\ell}=C_{ij\ell}+N_{ij\ell}$.
The quantity $d_\ell^{\rm mix}(\bm\Theta)$ in \autoref{eq:like} is constructed from $d_\ell^{\rm th}(\bm\Theta)$ by replacing, one after each other, the theoretical spectra with the corresponding observational ones \citep[for details, see][]{2013JCAP...01..026A}. 

Note that \autoref{eq:like} allows one to compute the difference between the $\chi^2$ at each point in the parameter space and its minimum value, with $\Delta\chi^2$ vanishing when computed using the fiducial values of our free parameters. This is the quantity that we compute at each step of our MCMC, i.e. a constant rescaling of the $\chi^2$ by an offset, which therefore correctly samples both the peak and the shape of the posterior, as it does not change the dependency of the $\chi^2$ on the free parameters.

For currently available observations, which are not able to survey extremely large volumes of the Universe and therefore do not explore the ultra-large-scale regime, the approximated spectra generally mimic the true power spectrum. Thus, the approximations made do not significantly affect the results. However, we expect future surveys, such as the HI-galaxy redshift survey for which we generated the mock data set in \autoref{sec:survey}, to provide data at scales where lensing, RSD, relativistic effects, and the Limber approximation significantly impact the power spectra. Consequently, using the different approximations presented in \autoref{sec:approx} in fitting the models to the data will likely lead to shifts in the inferred cosmological parameters with respect to the fiducial values used to generate the data set. 
In this paper, we quantify the significance of these shifts, in units of $\sigma$, as
\begin{equation}\label{eq:shift}
    S(\Theta) = \frac{|\Theta-\Theta^{\rm fid}|}{\sigma_\Theta},
\end{equation}
where $\Theta$ is a generic parameter of the full set $\bm\Theta$ estimated in our analysis, $\sigma_\Theta$ is the Gaussian error we obtain on $\Theta$, and $\Theta^{\rm fid}$ is the fiducial value of $\Theta$ used to generate the mock data set.

We apply this pipeline to the models described in \autoref{sec:cases}, with the baseline $\Lambda$CDM model described by the set of five free parameters $\bm\Theta=\{\omega_{\rm b}, \omega_{\rm c}, h, A_{\rm s}, n_{\rm s}\}$. When analysing an extended model, we add one extra free parameter to this set: the dark energy equation of state $w$, the sum of the neutrino masses $\sum{m_\nu}$, or the local primordial non-Gaussianity parameter $f_{\rm NL}$. We adopt flat priors on all these parameters.

Note that here we consider an optimistic setting in which the linear galaxy bias $\bg(z)$ is perfectly known. Adding nuisance parameters accounting for the uncertainty on this function and marginalising over them would enlarge the errors on cosmological parameters and reduce the statistical significance of the shifts we find, but would not qualitatively change the effects we are interested in. Moreover, as we are interested in the largest scales, in our analysis we only consider the data up to the multipole $\ell=400$. Adding smaller scales to the analysis could reduce the significance of the shifts, but would not change our results qualitatively.

\subsection{Debiasing constraints on cosmological parameters}\label{sec:debiasing}
\begin{figure}
    \centering
    \includegraphics[width=0.48\textwidth]{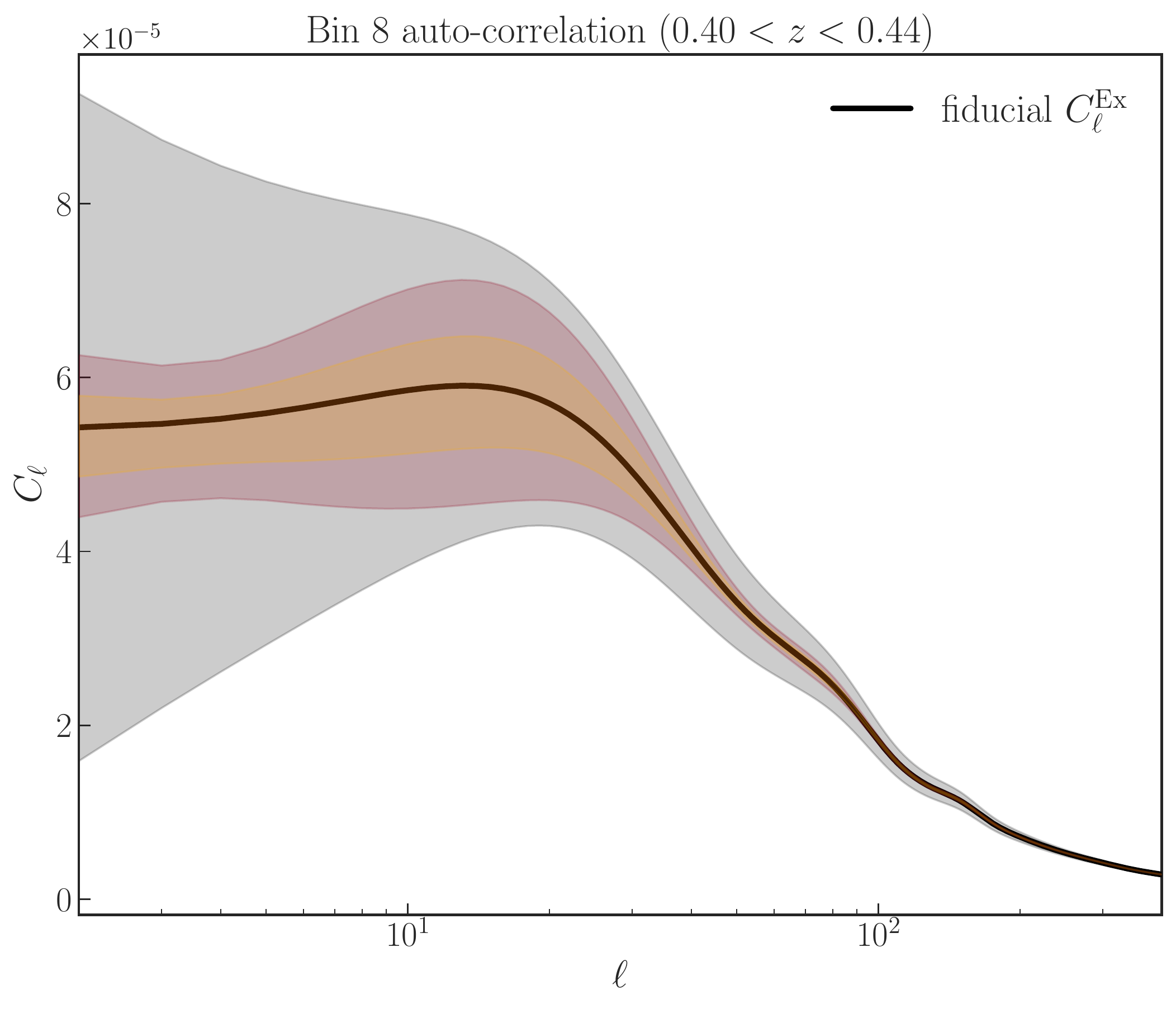}
    \caption{Effect of debiasing when different $\bm\Theta_0$ points in the parameter space are used to compute the debiasing term $\alpha(\bm\Theta^0)$. Here the auto-correlation in the eighth redshift bin is shown as a typical example. $\alpha(\bm\Theta^0)$ is computed at the fiducial set of parameters, $\bm\Theta^{\rm fid}$, and at $500$ other points in the parameter space, randomly sampled from a Gaussian distribution centred at $\bm\Theta^{\rm fid}$ with a variance of $10\%$ for each parameter. The black solid curve shows the fiducial $\cex(\bm\Theta^{\rm fid})$ spectrum, while the grey band shows the errors corresponding to the experimental setup considered in the paper. For each of the $500$ computed $\alpha(\bm\Theta^0)$, the debiasing term is applied to the $\capp (\bm\Theta^{\rm fid})$ spectrum. Assuming that the resulting spectra also follow a Gaussian distribution around the $\capp (\bm\Theta^{\rm fid})+\alpha(\bm\Theta^{\rm fid})$ spectra, the orange and red areas show the $1\sigma$ and $2\sigma$ uncertainty regions, respectively.}
    \label{fig:corr_ratio}
\end{figure}

As we will show in Section~\ref{sec:results}, using approximated spectra, $\capp$, in the MCMC analysis results in significant shifts on cosmological parameters. To mitigate this, we propose a method for debiasing the parameter estimates while still allowing for the use of the quickly computed $\capp$. This method is based on adding a correction to the $C_\ell$'s used in the likelihood evaluation as
\begin{equation}\label{eq:corrected_cl}
    \capp(\bm\Theta) \rightarrow \capp(\bm\Theta) + \left[\cex(\bm\Theta^0)-\capp(\bm\Theta^0)\right],
\end{equation}
where $\bm\Theta^0$ refers to a specific set of the cosmological parameters. We define the debiasing term $\alpha(\bm\Theta^0)$ as
\begin{equation}\label{eq:correction_factor_alpha}
    \alpha(\bm\Theta^0) \equiv  \cex(\bm\Theta^0)-\capp(\bm\Theta^0)\,.
\end{equation}
We use $\bm\Theta^0 = \bm\Theta^{\rm fid}$ for most of the results shown below, but discuss in \autoref{sec:mle_finding} how we can use a maximum likelihood estimate of $\bm\Theta^0$ when working with actual data for which $\bm\Theta^{\rm fid}$ does not exist.
In \autoref{fig:corr_ratio} we show the dependence of this debiasing method on the choice of $\bm\Theta^0$; we compute the debiasing term at $\bm\Theta^{\rm fid}$ and at $500$ other points of the parameter space, randomly sampled from a Gaussian distribution centred at $\bm\Theta^{\rm fid}$ with a variance on each parameter corresponding to $10\%$ of its fiducial value. These debiasing terms are then applied to $\capp(\bm\Theta^{\rm fid})$. Assuming that the resulting spectra also follow a Gaussian distribution around the $\capp+\alpha(\bm\Theta^{\rm fid})$ spectra, we show the corresponding $1\sigma$ and $2\sigma$ uncertainty regions. The figure shows that although the results we present below are based on computing $\alpha(\bm\Theta^0)$ using $\bm\Theta^{\rm fid}$, which would not be known in the case of actual data, our results would also hold for other choices of $\bm\Theta^0$ if they were reasonably close to $\bm\Theta^{\rm fid}$. This method of debiasing cosmological parameter estimates works precisely because the debiasing term $\alpha(\bm\Theta^0)$ is not strongly dependent on the choice of $\bm\Theta^0$ and can therefore account for the differences between the exact and approximated spectra over the full range of parameter space that the MCMC explores. Since $\alpha(\bm\Theta^0)$ only needs to be computed at a single set of parameter values, rather than each step in the MCMC, it allows one to obtain unbiased results without being computationally expensive, unlike using $\cex$ which makes the analysis unfeasible.

We therefore use, at each sampled point $\bm\Theta$, the $\chi^2$ expression of \autoref{eq:like}, but with the substitution
\begin{equation}
    \tilde{C}_{ij\ell}^{\rm Ap}(\bm\Theta)\rightarrow\tilde{C}_{ij\ell}^{\rm Ap}(\bm\Theta)+\alpha(\bm\Theta^0)\,.
\end{equation}

\subsection{Debiasing with maximum likelihood}\label{sec:mle_finding}

While in this paper we work with mock data sets, and therefore $\bm\Theta^{\rm fid}$ is known, this will not be the case when analysing real data. In order to use our approach in a realistic setting, we need to find a point in the parameter space that approximates the fiducial cosmology, which corresponds to the peak of the multivariate posterior probability distribution for the parameters. This can be achieved by analysing the mock data set built with $\cex$ using the correct theoretical predictions, but without attempting to reconstruct the full shape of the posterior distribution. One can use maximisation methods to find the peak of the distribution, and since these methods only aim to find the maximum likelihood (or best-fit) point in the parameter space, they require a significantly smaller number of iterations with respect to MCMC methods. 

Here, we use the maximisation pipeline of \texttt{Cobaya}, which relies on the \texttt{BOBYQA} algorithm \citep{2018arXiv180400154C,2018arXiv181211343C}, to fit the $\cex$ spectra to our mock data set, and we find the maximum likelihood parameter set presented in \autoref{tab:mle_results}. The maximum likelihood point ($\bm\Theta^{\rm peak}$) found with this method is very close to the actual fiducial point used to generate the data set and would therefore be suitable for computing the debiasing term $\alpha$ (see \autoref{sec:debiasing}). Although we use the fiducial parameter set $\bm\Theta^{\rm fid}$ to compute the debiasing term in the rest of this paper, we have verified that there would be no significant changes in our results if $\bm\Theta^{\rm peak}$ were chosen instead (see \autoref{sec:resLCDM}).

We find that the maximisation approach is much less computationally expensive compared to running a full MCMC with the exact spectra. A single iteration of our likelihood code using the exact spectra takes $\sim 150$ seconds (compared to $\sim 5$ seconds with the approximated spectra). The number of accepted iterations before reaching convergence \footnote{The convergence criteria used by \texttt{Cobaya} are that the Gelman-Rubin $R$-1 on the means be $<0.01$ and that on the standard deviations be $<0.2$.} with the approximated spectra is ~37,500. If we assume this to be the minimum number of iterations needed, the MCMC with the exact spectra would take at least 65 days (and likely much longer when taking into account the rejected steps). In contrast, the likelihood maximisation took under one month running on a workstation with many other background processes, and the MCMC run that followed (with the approximated spectra) took only 2 days, demonstrating the computational feasibility of our approach.

We want to stress, however, that this minimisation approach might fail should the posterior distribution be complicated; the presence of multiple peaks or very flat posteriors might bias the estimate of the maximum likelihood point in the parameter space and therefore possibly hinder the feasibility of this approach.

\begin{table}
\centering
\begin{tabular}{|c|c|c|}
\hline
$\Theta$                    & Fiducial value  & ML (or peak) value\\
\hline
$\omega_{\rm b}$          & $0.022445$   & $0.022485$ \\
$\omega_{\rm c}$          & $0.1206$     & $0.1209$ \\
$h$                       & $0.67$       & $0.67$ \\
$A_{\rm s}\times10^{-9}$  & $2.12605$    & $2.11$ \\
$n_{\rm s}$               & $0.96$       & $0.96$ \\
\hline
\end{tabular}
\caption{Maximum likelihood (ML) parameter set obtained by minimising the $\chi^2$ when $\cex$ is used to fit the data set described in Section~\ref{sec:survey}. The values are obtained through the \texttt{BOBYQA} minimisation algorithm implemented in \texttt{Cobaya}.}\label{tab:mle_results}
\end{table}

\section{Results and discussion}\label{sec:results}
In this section, we present the results of our analysis, highlighting how neglecting effects that are relevant at very large scales can result in significant biases in the estimation of cosmological parameters, potentially leading to false detections of non-standard physics. We split our results in two subsections, the first focusing on $\Lambda$CDM and its simple extensions $\Lambda$CDM+$m_\nu$ and $w$CDM, and the second discussing the results obtained when a scale-dependent bias generated by primordial non-Gaussianity is included in the analysis. All the MCMC samples obtained using the methodology described in the previous section are analyzed using the public code \texttt{GetDist}\footnote{\url{https://github.com/cmbant/getdist}} \citep{Lewis:2019xzd}.

\subsection{${\bf \Lambda}$CDM and its simple extensions}\label{sec:resLCDM}

\begin{table*}
\centering
\resizebox{\textwidth}{!}{
\renewcommand{\arraystretch}{1.5}
\begin{tabular}{ccccccccccr}
    \hline
     & & & \multicolumn{7}{c}{Cosmological parameters} & $\Delta\chi^2_{\rm min}$\\
    \cline{4-10}
     & & & $\omega_{\rm b}$ & $\omega_{\rm c}$ & $h$ & $A_{\rm s}\times10^9$ & $n_{\rm s}$ & $\sum m_\nu\,\mathrm{[eV]}$ & $w$ & \\
    \hline
    \hline
     & & $\Theta^{\rm fid}$ & $0.22445$ & $0.1206$ & $0.67$ & $2.12605$ & $0.96$ & $0.06$ & $-1$ & \\
    \hline
    \multirow{6}{*}{\rotatebox{90}{Biased results}} & \multirow{2}{*}{\rotatebox{90}{$\Lambda$CDM}} & $\Theta$ & $0.0163^{+0.0016}_{-0.0018}$ & $0.1098\pm0.0046$ & $0.616\pm0.018$ & $2.176^{+0.068}_{-0.081}$ & $0.9948\pm0.0071$ & $-$ & $-$ & \multirow{2}{*}{4972} \\
     & & $S(\Theta)\,[\sigma]$ & $3.6$ & $2.3$ & $3.0$ & $0.7$ & $4.9$ & $-$ & $-$ & \\
     \cline{2-11}
     & \multirow{2}{*}{\rotatebox{90}{$+m_\nu$}}& $\Theta$ & $0.0160^{+0.0015}_{-0.0018}$ & $0.1143^{+0.0044}_{-0.0052}$ & $0.615^{+0.017}_{-0.019}$ & $2.290\pm0.082$ & $0.9872\pm0.0074$ & $0.327\pm0.043$ & $-$ & \multirow{2}{*}{4925} \\
     & & $S(\Theta)\,[\sigma]$ & $3.8$ & $1.3$ & $3.0$ & $2.0$ & $3.7$ & $6.2$ & $-$ & \\
     \cline{2-11}
     & \multirow{2}{*}{\rotatebox{90}{$w$CDM}}& $\Theta$ & $0.0156^{+0.0013}_{-0.0016}$ & $0.1014\pm0.0039$ & $0.600\pm0.015$ & $2.398^{+0.079}_{-0.088}$ & $1.0494\pm0.0096$ & $-$ & $-0.886\pm0.013$ & \multirow{2}{*}{4912} \\
     & & $S(\Theta)\,[\sigma]$ & $4.7$ & $4.9$ & $4.7$ & $3.2$ & $9.3$ & $-$ & $8.7$ & \\
    \hline
    \multirow{6}{*}{\rotatebox{90}{Debiased results}} & \multirow{2}{*}{\rotatebox{90}{$\Lambda$CDM}} & $\Theta$ & $0.0233^{+0.0021}_{-0.0031}$ & $0.1227^{+0.0053}_{-0.0074}$ & $0.677^{+0.019}_{-0.026}$ & $2.104\pm0.078$ & $0.9573^{+0.0088}_{-0.0074}$ & $-$ & $-$ & \multirow{2}{*}{4.02} \\
     & & $S(\Theta)\,[\sigma]$ & $0.31$ & $0.33$ & $0.30$ & $0.28$ & $0.34$ & $-$ & $-$ & \\
     \cline{2-11}
     & \multirow{2}{*}{\rotatebox{90}{$+m_\nu$}}& $\Theta$ & $0.0235^{+0.0022}_{-0.0032}$ & $0.1238^{+0.0055}_{-0.0078}$ & $0.679^{+0.020}_{-0.027}$ & $2.113\pm0.091$ & $0.9563^{+0.0094}_{-0.0080}$ & $<0.115$ & $-$ & \multirow{2}{*}{4.11} \\
     & & $S(\Theta)\,[\sigma]$ & $0.39$ & $0.46$ & $0.39$ & $0.15$ & $0.41$ & $0.50$ & $-$ & \\
     \cline{2-11}
     & \multirow{2}{*}{\rotatebox{90}{$w$CDM}}& $\Theta$ & $0.0233^{+0.0021}_{-0.0032}$ & $0.1228^{+0.0054}_{-0.0078}$ & $0.677^{+0.019}_{-0.027}$ & $2.103\pm0.087$ & $0.957^{+0.012}_{-0.010}$ & $-$ & $-1.001^{+0.015}_{-0.013}$ & \multirow{2}{*}{0.07} \\
     & & $S(\Theta)\,[\sigma]$ & $0.30$ & $0.31$ & $0.29$ & $0.27$ & $0.28$ & $-$ & $0.07$ & \\
\end{tabular}
\renewcommand{\arraystretch}{1}
}
\caption{{\it Top table:} marginalised constraints on the sampled parameters $\bm\Theta$ and values of the shift estimator $S(\Theta)$ obtained by analysing the fiducial data set with the approximated $\capp$ spectra for the standard $\Lambda$CDM model and its simple extensions, $\Lambda$CDM+$m_\nu$ and $w$CDM, considered in the present work. The last column of the table shows the minimum $\Delta\chi^2$ values obtained for the different cosmologies, which, given \autoref{eq:like}, should vanish for an unbiased analysis. {\it Bottom table:} same as the top table, but applying the debiasing term $\alpha(\bm\Theta^{\rm fid})$ to the theoretical predictions that are then compared with the data.}\label{tab:extshift}
\end{table*}

\begin{figure*}
    \centering
    \begin{tabular}{ccc}
    \includegraphics[width=0.3\textwidth]{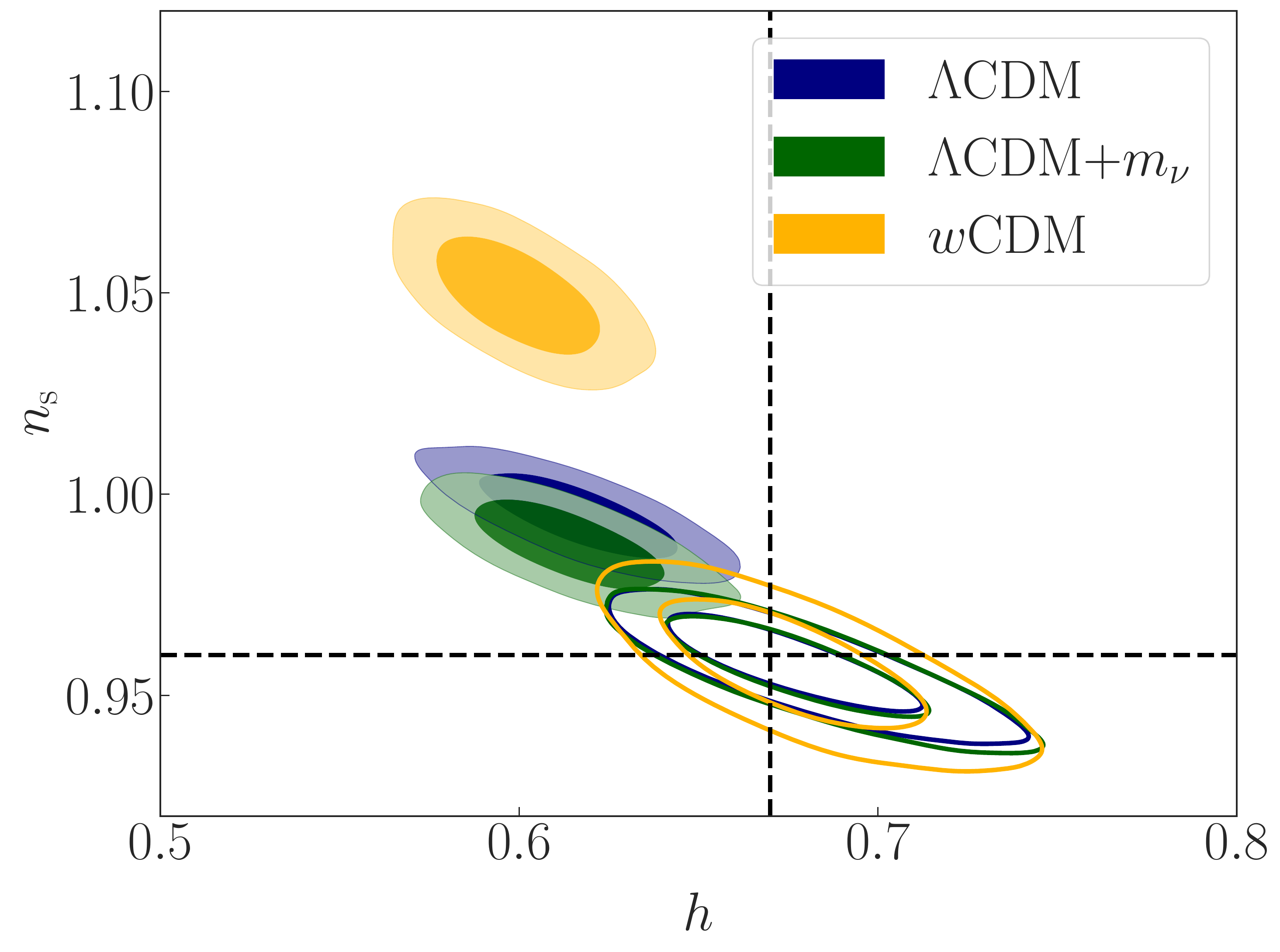} &
    \includegraphics[width=0.3\textwidth]{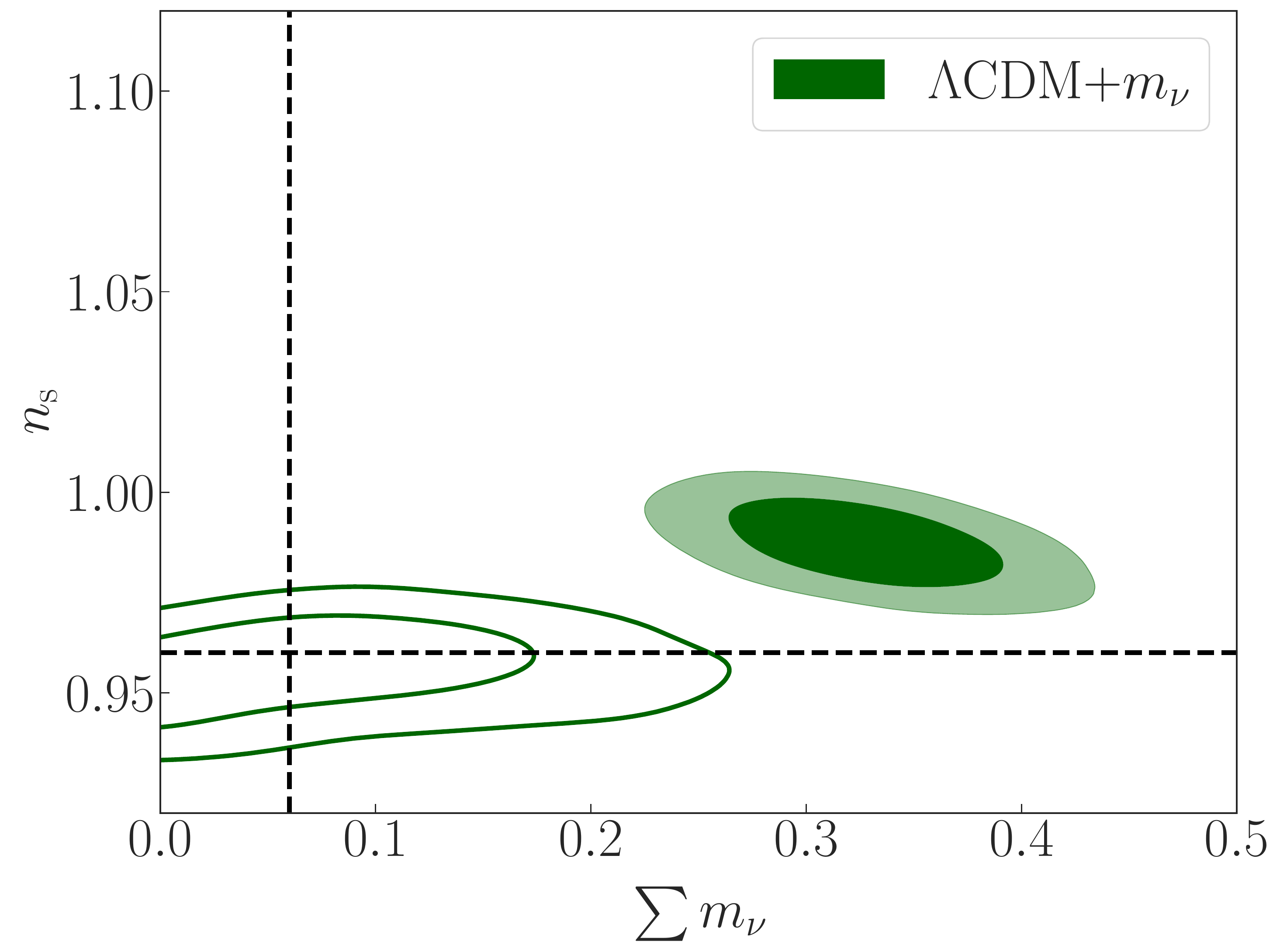} &
    \includegraphics[width=0.3\textwidth]{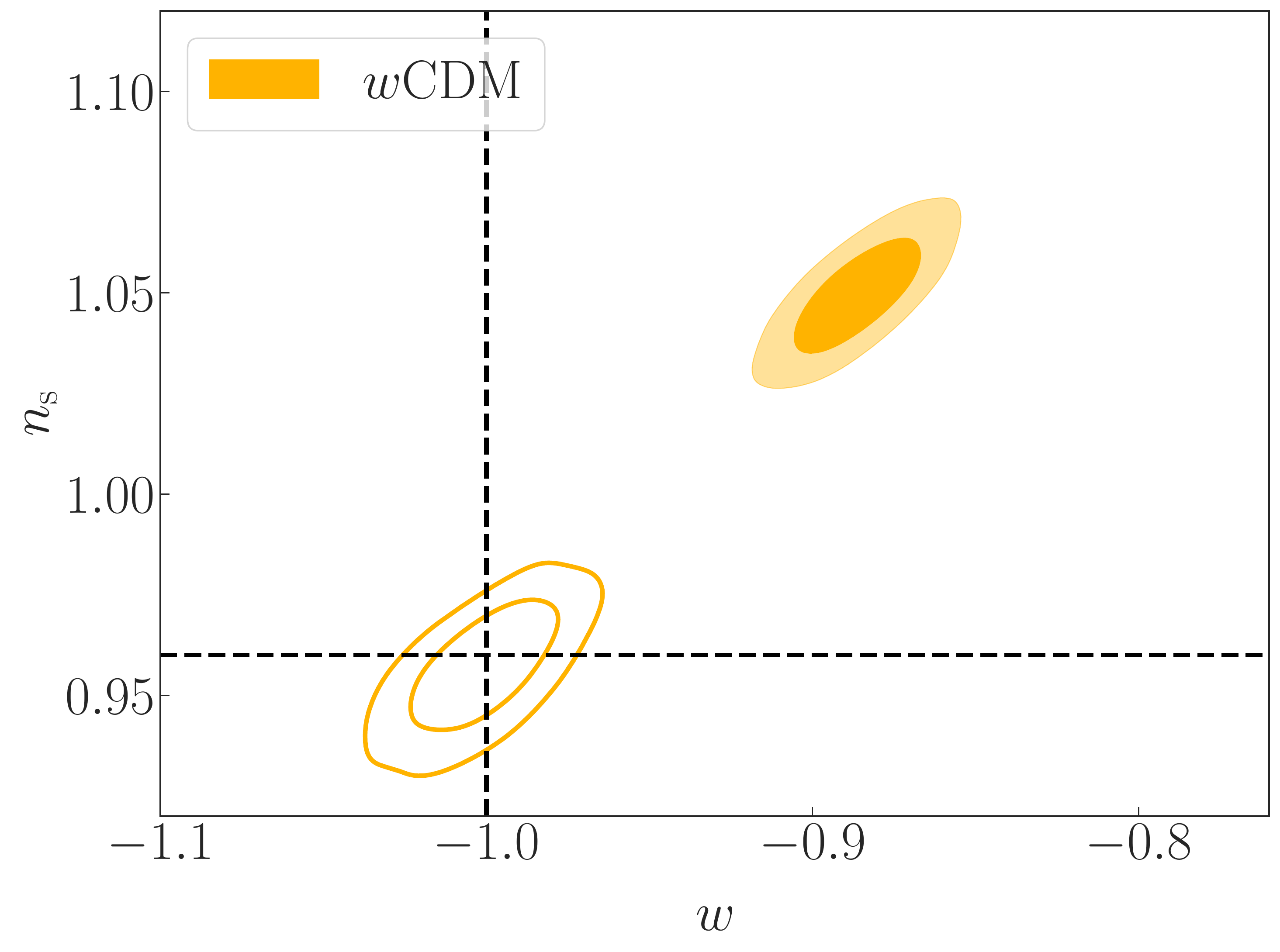} \\
    \end{tabular}
    \caption{$68\%$ and $95\%$ confidence level contours obtained by fitting the approximated $\capp$ spectra to the data set built using the exact $\cex$ spectra (colour-filled contours). The violet contours show the result when a $\Lambda$CDM model is assumed, while the green and orange contours correspond instead to $\Lambda$CDM+$m_\nu$ and $w$CDM cosmologies, respectively. The empty contours show the results of the analysis when the debiasing term described in \autoref{sec:debiasing} is included. The black dashed lines show the fiducial values of the cosmological parameters.}
    \label{fig:biasedres}
\end{figure*}

In \autoref{tab:extshift}, we present the results obtained by analysing our mock data set, generated with $\cex$ spectra for a $\Lambda$CDM fiducial cosmology, using $\capp$ spectra for the three assumed cosmologies $\Lambda$CDM, $\Lambda$CDM+$m_\nu$ and $w$CDM. In the first case, we find that the obtained constraints on cosmological parameters are significantly shifted with respect to their fiducial values, despite using for the analysis the same cosmological model as the one assumed in generating the mock data set. With the exception of $A_{\rm s}$, which affects the amplitude of the spectra, the other parameters are all shifted by more than $2\sigma$, with $n_{\rm s}$ being the most affected parameter ($S=4.9\sigma)$, as a result of using approximations to achieve a reasonable computation time for the MCMC analysis. When we allow for simple extensions of $\Lambda$CDM, we see that such an effect leads to significant false detections of departures from the standard model. With the sum of the neutrino masses $\sum{m_\nu}$ added as an extra free parameter, we indeed find a significant detection of a non-vanishing value, where $\sum{m_\nu}=0$~eV is excluded with more than $6\sigma$ significance and the estimated value is shifted from the fiducial minimal value $\sum{m_\nu}=0.06$~eV by $S=6.2\sigma$; this implies that an analysis of data sensitive to large-scale effects would provide a false detection of the neutrino masses if one used the approximations considered here. The same effect can be seen if one allows for dark energy with an equation of state parameter that deviates from the cosmological constant value ($w=-1$). In this case, the free parameter $w$ is shifted from the fiducial value by $S=8.7\sigma$, resulting in a significant detection of a non-standard behaviour, which is driven only by the use of the approximated $\capp$ in the parameter estimation pipeline. Also in these extended cases, the estimated values of the standard cosmological parameters are shifted with respect to the fiducial ones. This highlights how these simple extensions alone are not able to mimic the $\cex$ spectra, as shifts in the values of the standard parameters are also necessary for fitting the $\cex$ to the data when $\capp$ are being used. The new degeneracies introduced by extensions of the $\Lambda$CDM model explain the changes in values of $S$ with respect to the standard model.

We also show in \autoref{tab:extshift} the minimum value of the $\Delta\chi^2$ found by analysing the posterior distribution reconstructed with the MCMC ($\Delta\chi^2_{\rm min}$). The decrease in the values of $\Delta\chi^2_{\rm min}$ for the extended models with respect to $\Lambda$CDM shows that a false detection of the extensions allows the approximated spectra to be in better agreement with the data. However, given that we expect from \autoref{eq:like} to obtain a $\Delta\chi^2_{\rm min}$ close to zero if the theoretical spectra match the data, the values shown in \autoref{tab:extshift} highlight how even these significant shifts in the $\capp$ spectra are not able to reproduce the cosmology used to generate the data. Notice that here we are not suggesting that the reduction in $\Delta\chi^2_{\rm min}$ is pointing towards a statistical preference for one model over the other; such a comparison would require using Bayesian model comparison techniques also accounting for the number of free parameters of a given model. Moreover, the $\Delta\chi^2_{\rm min}$ values are estimates that might be slightly different from the real minimum value, as it is not guaranteed that the MCMC is able to perfectly sample the peak point in the parameter space. Thus, with such small differences between different models, a more accurate computation of $\Delta\chi^2_{\rm min}$ would be needed if one wanted to perform model comparison.

In \autoref{fig:biasedres}, we show the $68\%$ and $95\%$ confidence level contours on a few representative parameters for the cases described above. The colour-filled contours show the results of the analysis performed with $\capp$, highlighting the deviation of the estimated values of the parameters from the fiducial values (shown with black dashed lines). The empty contours instead show the results obtained when the debiasing term described in \autoref{sec:debiasing} is added to the spectra, which are then compared to the mock data set. These results show how the method we propose is able to debias the results and how it allows us to recover the correct values for the parameters, for both the standard $\Lambda$CDM cosmology and its extensions, thus avoiding false detections of non-standard cosmologies and improving the goodness of fit with a $\chi^2$ now of $\mathcal{O}(1)$.

\begin{figure}
    \centering
    \includegraphics[width=0.45\textwidth]{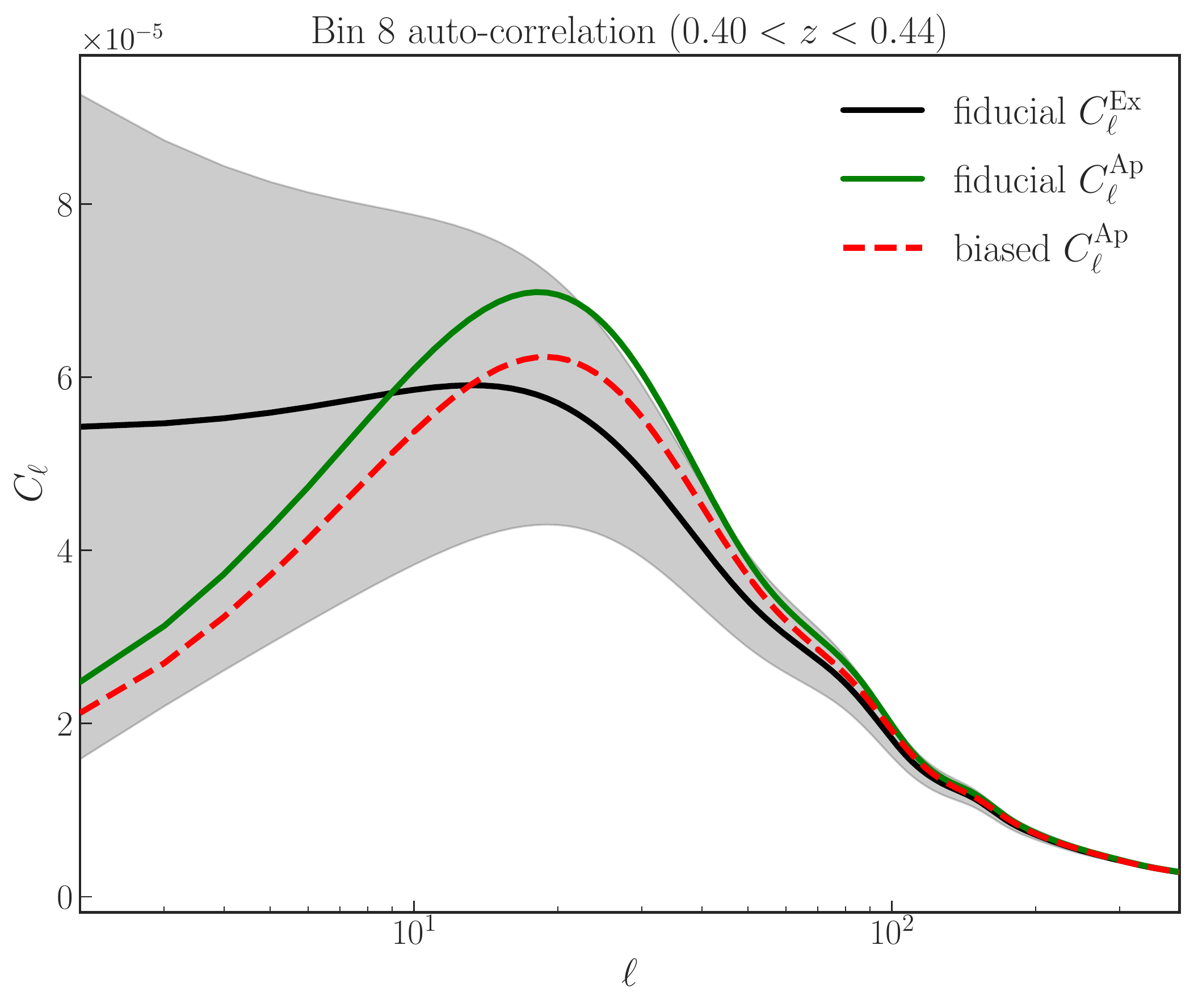}
    \caption{Angular power spectra for the eighth redshift bin auto-correlation in a $\Lambda$CDM cosmology using the exact $\cex$ (black solid curve) and the approximated $\capp$ (green solid curve) obtained assuming the fiducial values for the cosmological parameters. The red dashed curve shows the $\capp$ obtained for the biased parameter estimation of \autoref{tab:extshift}. The grey area shows the errors corresponding to the experimental setup used throughout the paper.}
    \label{fig:biased_spectra}
\end{figure}

In order to see in more detail the biasing effect of the approximations included in the $\capp$, we show in \autoref{fig:biased_spectra} the impact of the biases on the angular power spectra for a representative redshift bin auto-correlation, highlighting how the approximated $\capp$ spectrum (green) significantly departs from the expected $\cex$ spectrum (black) when the fiducial values of the cosmological parameters are used to obtain both. We also include, with a red dashed curve, the $\capp$ spectrum obtained using the biased values of the cosmological parameters reported in \autoref{tab:extshift}, showing how in this case the $\capp$ at the shifted best-fit cosmology are better able to reproduce the fiducial $\cex$, thus producing a better fit to the data. 

\begin{figure}
    \centering
    \includegraphics[width=0.48\textwidth]{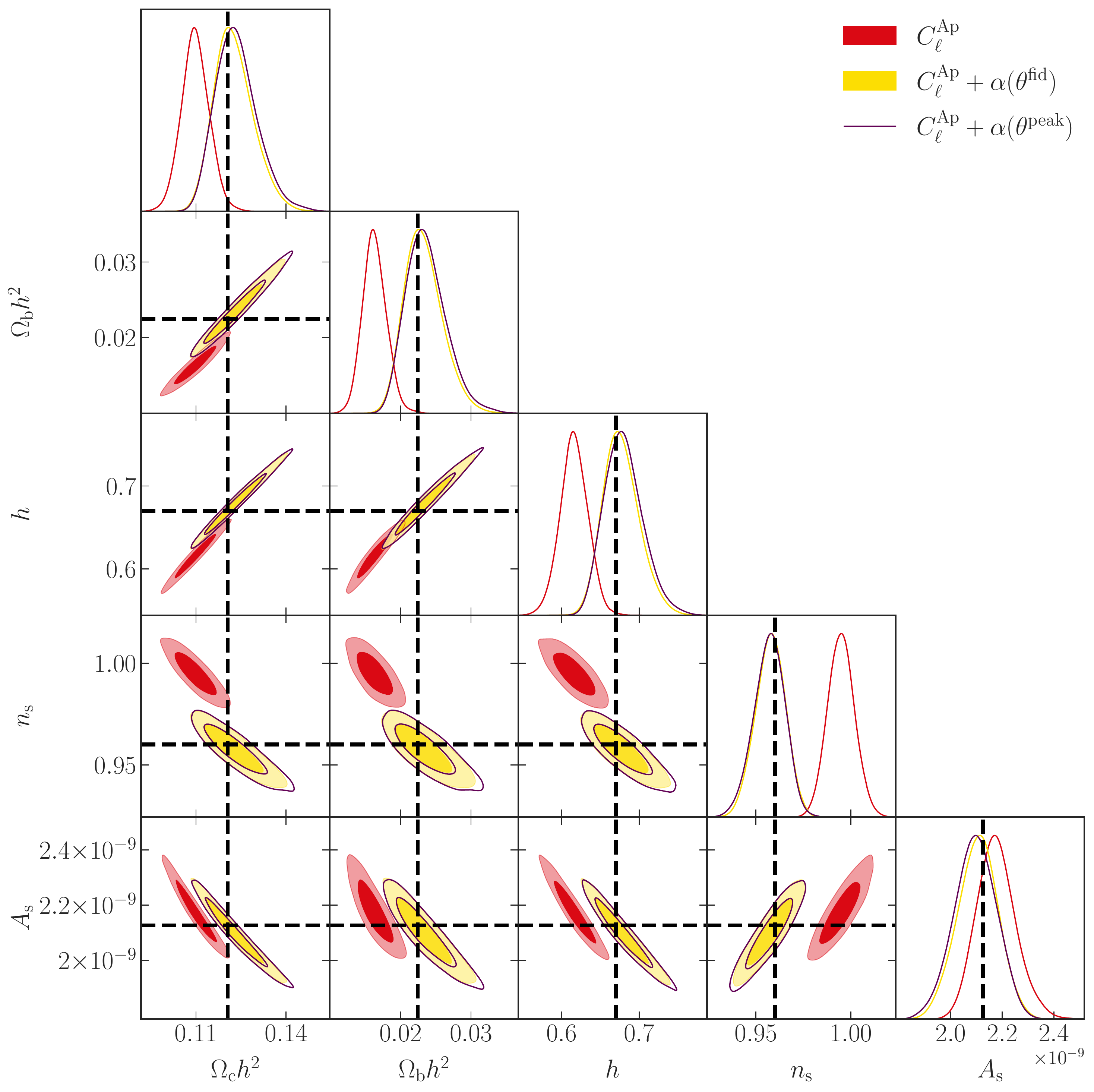}
    \caption{$68\%$ and $95\%$ confidence level contours, as well as one-dimensional marginalised posterior probability distribution functions, obtained by fitting the $\Lambda$CDM model to the mock data set. The red, filled contours correspond to the analysis where the theoretically predicted approximated spectra $\capp$ are used to fit the model to the mock data set. The yellow, filled contours show the results obtained when the debiasing term $\alpha(\bm\Theta^{\rm fid})$ is included, and the purple, empty contours correspond to the debiasing term $\alpha(\bm\Theta^{\rm peak})$ computed at the estimated maximum likelihood point $\bm\Theta^{\rm peak}$.}
    \label{fig:triangle_LCDM}
\end{figure}

While in \autoref{fig:biasedres} we only show a subset of the free parameters of our models, the debiasing procedure is effective for all cosmological parameters. In \autoref{fig:triangle_LCDM}, we show the constraints obtained on all the free parameters of our $\Lambda$CDM analysis, obtained by both comparing the $\capp$ to the mock data set (red, filled contours) and applying the debiasing method of \autoref{sec:debiasing}, using the debiasing term computed at both the fiducial values, $\alpha(\bm\Theta^{\rm fid})$ (yellow, filled contours), and the peak values found in \autoref{sec:mle_finding}, $\alpha(\bm\Theta^{\rm peak})$ (purple, empty contours). We notice how in the first case all the parameters are shifted with respect to their expected values, with the most significant shifts on $n_s$, $\omega_{\rm b}$ and $h$, while when we apply the debiasing approach the fiducial values are recovered for all the parameters, with no significant differences between the two cases of $\alpha(\bm\Theta^{\rm fid})$ and $\alpha(\bm\Theta^{\rm peak})$. Even though the results shown in \autoref{fig:triangle_LCDM} correspond to the $\Lambda$CDM model, they are qualitatively similar for all the considered cosmologies.

The posterior probability distributions of the parameters recovered after debiasing the MCMC results do not necessarily coincide with those that would be obtained by a full analysis. We can, however, consider these as reasonable estimates, as \autoref{fig:corr_ratio} shows that the debiasing term does not depend strongly on the $\bm\Theta^0$ point at which it is computed, as long as $\alpha(\bm\Theta^0)$ is sufficiently close to $\bm\Theta^{\rm fid}$. Thus, rather than computing $\alpha(\bm\Theta^0)$ at each point in the parameter space, we can approximate $\alpha(\bm\Theta^0)$ with $\alpha(\bm\Theta^{\rm peak})$ (or $\alpha(\bm\Theta^{\rm fid})$ in the case of the results shown here). This applies only in the vicinity of the peak of the distribution, and the estimation of the tails suffers from an error that propagates into the confidence intervals shown in \autoref{fig:biasedres} and \autoref{fig:triangle_LCDM}. We leave a quantification of this error for future work.

\subsection{Primordial non-Gaussianity}\label{sec:resfNL}

\begin{table*}
\centering
\resizebox{\textwidth}{!}{
\renewcommand{\arraystretch}{1.5}
\begin{tabular}{cccccccccr}
    \hline
     & & & \multicolumn{6}{c}{Cosmological parameters} & $\Delta\chi^2_{\rm min}$\\
    \cline{3-9}
     & & & $\omega_{\rm b}$ & $\omega_{\rm c}$ & $h$ & $A_{\rm s}\times10^9$ & $n_{\rm s}$ & $f_{\rm NL}$ & \\
    \hline
    \hline
     & & $\Theta^{\rm fid}$ & $0.22445$ & $0.1206$ & $0.67$ & $2.12605$ & $0.96$ & $0$ & \\
    \hline
    \multirow{6}{*}{\rotatebox{90}{Biased results}} & \multirow{2}{*}{\rotatebox{90}{baseline}} & $\Theta$ & $0.0164^{+0.0016}_{-0.0019}$ & $0.1101^{+0.0044}_{-0.0050}$ & $0.617\pm0.018$ & $2.170\pm0.077$ & $0.9945\pm0.0070$ & $-0.8\pm3.9$ & \multirow{2}{*}{4985} \\
     & & $S(\Theta)\,[\sigma]$ & $3.4$ & $2.2$ & $2.9$ & $0.6$ & $4.9$ & $0.2$ & \\
     \cline{2-10}
     & \multirow{2}{*}{\rotatebox{90}{$z$ cut}}& $\Theta$ & $0.0219^{+0.0024}_{-0.0035}$ & $0.1300^{+0.0064}_{-0.0089}$ & $0.679^{+0.023}_{-0.031}$ & $1.839\pm0.084$ & $0.9955^{+0.0093}_{-0.0081}$ & $-85^{+13}_{-12}$ & \multirow{2}{*}{1175} \\
     & & $S(\Theta)\,[\sigma]$ & $0.2$ & $1.2$ & $0.3$ & $3.4$ & $4.0$ & $6.9$ & \\
     \cline{2-10}
     & \multirow{2}{*}{\rotatebox{90}{$p=0.5$}}& $\Theta$ & $0.0169^{+0.0017}_{-0.0021}$ & $0.1143^{+0.0048}_{-0.0056}$ & $0.620\pm0.020$ & $2.069\pm0.077$ & $1.0047\pm0.0075$ & $66.5\pm7.2$ & \multirow{2}{*}{4755} \\
     & & $S(\Theta)\,[\sigma]$ & $2.9$ & $1.2$ & $2.6$ & $0.7$ & $6.0$ & $9.2$ & \\
    \hline
    \multirow{6}{*}{\rotatebox{90}{Debiased results}} & \multirow{2}{*}{\rotatebox{90}{baseline}} & $\Theta$ & $0.0235^{+0.0022}_{-0.0033}$ & $0.1230^{+0.0054}_{-0.0079}$ & $0.678^{+0.020}_{-0.028}$ & $2.101^{+0.087}_{-0.077}$ & $0.9573^{+0.0086}_{-0.0076}$ & $0\pm10$ & \multirow{2}{*}{17.8} \\
     & & $S(\Theta)\,[\sigma]$ & $0.34$ & $0.35$ & $0.33$ & $0.30$ & $0.33$ & $0.04$ & \\
     \cline{2-10}
     & \multirow{2}{*}{\rotatebox{90}{$z$ cut}}& $\Theta$ & $0.0235^{+0.0025}_{-0.0038}$ & $0.1232^{+0.0063}_{-0.0091}$ & $0.679^{+0.023}_{-0.032}$ & $2.101\pm0.096$ & $0.9573^{+0.0094}_{-0.0083}$ & $-1\pm12$ & \multirow{2}{*}{17.9} \\
     & & $S(\Theta)\,[\sigma]$ & $0.31$ & $0.32$ & $0.30$ & $0.26$ & $0.30$ & $0.11$ & \\
     \cline{2-10}
     & \multirow{2}{*}{\rotatebox{90}{$p=0.5$}}& $\Theta$ & $0.0234^{+0.0022}_{-0.0033}$ & $0.1229^{+0.0055}_{-0.0080}$ & $0.678^{+0.020}_{-0.028}$ & $2.103^{+0.088}_{-0.079}$ & $0.9575^{+0.0088}_{-0.0077}$ & $0.1\pm5.5$ & \multirow{2}{*}{17.9} \\
     & & $S(\Theta)\,[\sigma]$ & $0.31$ & $0.32$ & $0.30$ & $0.27$ & $0.30$ & $0.03$ & \\
\end{tabular}
\renewcommand{\arraystretch}{1}
}
\caption{{\it Top table:} marginalised constraints on the sampled parameters $\bm\Theta$ and values of the shift estimator $S(\Theta)$ obtained by analysing the fiducial data set with the approximated $\capp$ spectra for the three cases of $\Lambda$CDM$+f_{\rm NL}$ considered in the present work. {\it Bottom table:} same as the top one, but applying the debiasing term $\alpha(\bm\Theta^{\rm fid})$ to the theoretical predictions that are then compared with the data.}\label{tab:fNLshift}
\end{table*}

In this subsection, we focus on the results when $f_{\rm NL}$ is included as a free parameter, thus allowing for a non-vanishing local primordial non-Gaussianity; this affects the galaxy clustering spectra through the scale-dependent bias as described in \autoref{sec:fnl}. As a first case, we use the same experimental setup we used in \autoref{sec:resLCDM}, and use the standard expression of \autoref{eq:scaledepbias} for our theoretical predictions for the scale-dependent bias. In this case, which we refer to as `baseline', when we analyse the mock data set using the approximated $\capp$ we find results that are similar to the $\Lambda$CDM case of \autoref{sec:resLCDM}, with approximately the same shifts for the standard parameters and no bias for $f_{\rm NL}$ (see \autoref{tab:fNLshift}). This may seem to be a surprising result, as the impact of $f_{\rm NL}$ on the theoretical predictions is significant at very large scales (see \autoref{fig:theory}), where the approximations included in $\capp$ fail. One would therefore expect that a biased value for this parameter would help with fitting the $\cex$ spectra of the mock data set, and that a false non-vanishing $f_{\rm NL}$ would be detected. However, given \autoref{eq:scaledepbias} that we rely upon, the scale-dependent bias depends not only on $f_{\rm NL}$, but also on the $\bg-1$ factor. As shown in \autoref{fig:specs} and discussed in \autoref{sec:fnl}, our choice of the linear galaxy bias implies that $\bg-1$ changes sign at $z\approx0.75$; the impact of $f_{\rm NL}$ on the $\capp$ spectra is therefore the opposite for the redshift bins beyond this redshift threshold with respect to the lower-redshift ones. Such an effect leads to a cancellation of the impact of the primordial non-Gaussianity on the goodness of fit, and therefore the standard case of $f_{\rm NL}=0$ is still preferred.

In order to ensure that this indeed is the reason for the lack of shift in the recovered $f_{\rm NL}$ value, we run our parameter estimation pipeline by removing the redshift bins above $z\approx0.75$. We refer to this case as `$z$ cut'. The results are shown in \autoref{tab:fNLshift}, where it can be seen how removing the higher-redshift bins eliminates the cancellation effect described above; now we find significant biases on $f_{\rm NL}$ and $A_{\rm s}$, with $S(f_{\rm NL})=6.9\sigma$ and $S(A_{\rm s})=3.4\sigma$, respectively, for the shifts with respect to the fiducial values. The shifts on the other free parameters are reduced with respect to the baseline case. The combined effect of $f_{\rm NL}$ and $A_{\rm s}$ allows the $\capp$ to fit the mock data set, as the global effect is boosting the power spectra at large scales.

\begin{figure}
    \centering
    \includegraphics[width=0.45\textwidth]{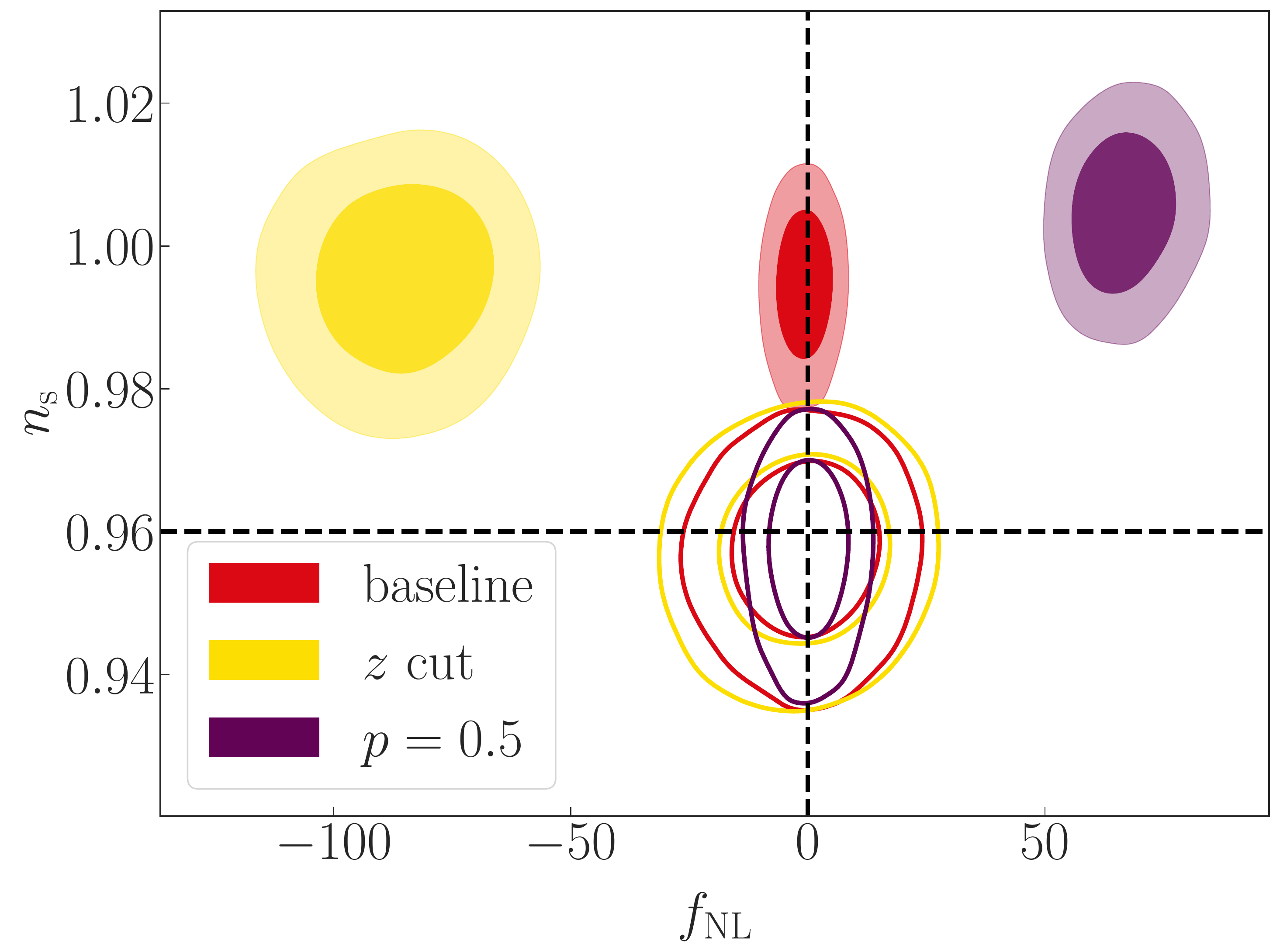}
    \caption{$68\%$ and $95\%$ confidence level contours obtained by fitting the approximated $\capp$ spectra, with a free $f_{\rm NL}$ parameter, to the data set built using the exact $\cex$ spectra (colour-filled contours). The red and yellow contours show the results obtained with the scale-dependent bias of \autoref{eq:scaledepbias}, with our baseline settings and with removing the last two redshift bins, respectively. 
    The violet contour shows instead the case where the scale-dependent bias is computed following \autoref{eq:scaledepbias-p} with $p=0.5$. The empty contours show the results of the analysis when the debiasing term described in \autoref{sec:debiasing} is included. The black dashed lines show the values of the fiducial cosmological parameters. }
    \label{fig:fNLbiasedres}
\end{figure}

On the other hand, as we discussed in \autoref{sec:fnl}, the modulating factor $\bg(z)-1$ in \autoref{eq:scaledepbias} is not the only possibility for describing the scale-dependent bias. We have repeated our analysis, following the more general \autoref{eq:scaledepbias-p}, by setting $p=0.5$, which ensures that the $\bg(z)-p$ factor does not change sign in our redshift range, given our choice of the linear galaxy bias. In the last two columns of \autoref{tab:fNLshift} we report the results we find in this case, where we see again a significant false detection of a non-vanishing $f_{\rm NL}$, with $S(f_{\rm NL})=9.2\sigma$, while the other parameters are less shifted from their fiducial values compared to the baseline case, with the exception of $n_{\rm s}$. In \autoref{fig:fNLbiasedres}, we also notice how the shift on $f_{\rm NL}$ has an opposite sign in this $p=0.5$ case with respect to the $z$ cut case, where the analysis prefer a negative value of $f_\mathrm{NL}$. This is due to the fact that the $\bg-p$ factor is now always positive, and one needs an $f_{\rm NL}>0$ in order to achieve the boost in the $\capp$ needed to fit the model to the mock data set.

Finally, we apply the debiasing procedure of \autoref{sec:debiasing} to the three cases described and show the results in \autoref{fig:fNLbiasedres}. As the figure shows, applying the debiasing correction allows us to recover a vanishing $f_{\rm NL}$. The debiased contours are different from each other here, which was not the case in \autoref{sec:resLCDM}; this is due to the different strategies applied to account for the effects of $f_{\rm NL}$ in our analysis.

\section{Conclusions}\label{sec:conclusions}

The continual improvement in galaxy surveys will soon unlock the largest scales in the sky for cosmological studies. While the expected angular correlations at smaller scales are well understood and efficiently modelled (up to the nonlinear regime), calculations of power spectra commonly make use of approximations aimed at reducing the computational efforts needed to obtain theoretical predictions of the spectra. This is a necessary requirement for such calculations if one wants to exploit MCMC methods for performing parameter estimation analyses. Such approximations, however, break down at very large scales, where effects including lensing, galaxy velocities and relativistic corrections become relevant.

In this paper, we have investigated the impact of approximations that neglect such large-scale effects on a parameter estimation analysis. We have produced a mock data set for a next-generation survey, with specifications based on those envisaged for the SKAO, that will be able to explore the angular correlation of galaxies at very large scales through the full treatment described in \autoref{sec:std}. We have then analysed this data set by applying the commonly used approximations described in \autoref{sec:approx}, where the large-scale corrections due to lensing, velocities and relativistic effects have been neglected, and the Limber approximation has been employed. We have found that this analysis produces significantly biased results, with parameter estimates being shifted up to $\sim 5\sigma$ when assuming a minimal 5-parameter $\Lambda$CDM cosmology, and with false detections of non-standard cosmologies when simple extensions of the standard model are considered.

We have also explored the impact of the approximations on a more complex extension of the $\Lambda$CDM model, where we have allowed for a non-vanishing local primordial non-Gaussianity by including $f_{\rm NL}$ as a free parameter in our analysis. This contributes a scale-dependent term to the galaxy bias which is relevant at large scales. We expected estimates of this parameter to be significantly biased, as a non-zero $f_{\rm NL}$ would help the approximated spectra to mimic those used in creating the data set. However, we have found that in our baseline setting, such an effect cannot be seen due to a cancellation between the low- and high-redshift bins. Given our choice of the linear galaxy bias (see \autoref{sec:survey}), the commonly used scale-dependent term changes sign  at $z\approx0.75$ and, therefore, the effect of a non-vanishing $f_{\rm NL}$ on the overall goodness of fit cancels out between low- and high-redshift bins. We have confirmed this explanation by cutting out all bins at $z>0.75$, and we have found, with this setting, a significant false detection of a non-vanishing and negative $f_{\rm NL}$. We have also performed our analysis for a case where the scale-dependent piece of the bias depends differently on the linear bias term (\autoref{eq:scaledepbias-p}). We have found in this case a $9.2\sigma$ shift in the estimated value of $f_{\rm NL}$, opposite in sign with respect to the previous case, highlighting how different modellings of the scale-dependent term can affect the final results.

In this work, not only have we assessed the impact of the approximations on the estimation of cosmological parameters, but we have also proposed a simple method to obtain debiased results that can approximate those that one would obtain by taking into account all the effects. We have described this method in \autoref{sec:debiasing} and pointed out how the computation of the debiasing term $\alpha({\bm\Theta^0})$ does not depend strongly on the choice of the parameter set $\bm\Theta^0$ where the computation is performed, as long as it is close to the true cosmology. Indeed our advantage in using this method relies on the fact that, in our forecasts, the fiducial cosmology has been known. However, we have pointed out that in a realistic setting, with an unknown fiducial cosmology, one could rely on minimisation algorithms to identify the best-fit point in the parameter space. Such a minimisation would be significantly less computationally expensive than a full parameter estimation pipeline and could therefore be performed using the exact spectra. We have tested the feasibility of such an approach, and we found in \autoref{sec:mle_finding} that the debiased cosmological parameter constraints found using an estimate for the peak of the multivariate distribution are almost exactly the same as those found using the fiducial point. Thus, this method can be applied to real data, where the fiducial point is unknown.

We have applied the debiasing method to all the cases we have investigated, and we have found that it indeed allows us to recover the expected values for the free parameters of our analyses. This method could therefore be used in real data analysis when unexpected detections of non-standard behaviour are seen. Additionally, while not providing a fully correct parameter estimation, our method allows one to obtain accurate values for cosmological parameters and estimates of their corresponding posterior probability distributions. While the recovered distributions are reasonable estimates of the ones obtained through a full analysis, we leave a quantitative assessment of the errors on their shapes for future work.

\section*{Acknowledgements}
We thank Michael Strauss for useful comments on a previous version of the manuscript. M.M. has received the support of a fellowship from `la Caixa' Foundation (ID 100010434), with fellowship code LCF/BQ/PI19/11690015, and the support of the Spanish Agencia Estatal de Investigacion through the grant `IFT Centro de Excelencia Severo Ochoa SEV-2016-0599'. R.D. acknowledges support from the Fulbright U.S.\ Student Program and the NSF Graduate Research Fellowship Program under Grant No.\ DGE-2039656. Any opinions, findings, and conclusions or recommendations expressed in this material are those of the authors and do not necessarily reflect the views of the National Science Foundation. Y.A. is supported by LabEx ENS-ICFP: ANR-10-LABX-0010/ANR-10-IDEX-0001-02 PSL*. S.C. acknowledges support from the `Departments of Excellence 2018-2022' Grant (L.\ 232/2016) awarded by the Italian Ministry of University and Research (\textsc{mur}). S.C. also acknowledges support by \textsc{mur} Rita Levi Montalcini project `\textsc{prometheus} -- Probing and Relating Observables with Multi-wavelength Experiments To Help Enlightening the Universe's Structure', for the early stages of this project, and from the `Ministero degli Affari Esteri della Cooperazione Internazionale (\textsc{maeci}) -- Direzione Generale per la Promozione del Sistema Paese Progetto di Grande Rilevanza ZA18GR02.

\section*{Data Availability}
The data underlying this article will be shared on reasonable request to the corresponding author.


\bibliographystyle{mnras}
\bibliography{bibliography} 

\bsp	
\label{lastpage}
\end{document}